\documentclass[%
 reprint,
 superscriptaddress,
 amsmath,amssymb,
 aps,
]{revtex4-2}

\usepackage{graphicx}
\usepackage{dcolumn}
\usepackage{bm}

\usepackage{braket}
\usepackage[qm]{qcircuit}
\usepackage[table]{xcolor}

\usepackage{hyperref}

\newcommand{\GR}[2]{\parbox{0.7cm}{\centering GR \\ \hspace{0pt} \\ $#1, #2 $}} 
\newcommand{\VZ}{\parbox{0.7cm}{\centering $Z$ \\ {\!\!\tiny(virtual)}}} 
\newcommand{\RZ}[1]{R_Z(#1)}
\newcommand{\rzfrac}[2]{#1/#2}

\begin{document}

\preprint{APS/123-QED}

\title{Fault-Tolerant Operation and Materials Science  \\
with Neutral Atom Logical Qubits}

\author{Matt. J. Bedalov}
\affiliation{Infleqtion (Boulder, CO, USA; Chicago, IL, USA; Louisville, CO, USA; Madison, WI, USA; Oxford, England, UK)}

\author{Matt Blakely}
\affiliation{Infleqtion (Boulder, CO, USA; Chicago, IL, USA; Louisville, CO, USA; Madison, WI, USA; Oxford, England, UK)}

\author{Peter. D. Buttler}
\affiliation{Infleqtion (Boulder, CO, USA; Chicago, IL, USA; Louisville, CO, USA; Madison, WI, USA; Oxford, England, UK)}

\author{Caitlin Carnahan}
\affiliation{Infleqtion (Boulder, CO, USA; Chicago, IL, USA; Louisville, CO, USA; Madison, WI, USA; Oxford, England, UK)}

\author{Frederic T. Chong}
\affiliation{Infleqtion (Boulder, CO, USA; Chicago, IL, USA; Louisville, CO, USA; Madison, WI, USA; Oxford, England, UK)}
\affiliation{Department of Computer Science, University of Chicago, Chicago, IL, USA}

\author{Woo Chang Chung}
\affiliation{Infleqtion (Boulder, CO, USA; Chicago, IL, USA; Louisville, CO, USA; Madison, WI, USA; Oxford, England, UK)}

\author{Dan C. Cole}
\affiliation{Infleqtion (Boulder, CO, USA; Chicago, IL, USA; Louisville, CO, USA; Madison, WI, USA; Oxford, England, UK)}

\author{Palash Goiporia}
\affiliation{Infleqtion (Boulder, CO, USA; Chicago, IL, USA; Louisville, CO, USA; Madison, WI, USA; Oxford, England, UK)}

\author{Pranav Gokhale}
\affiliation{Infleqtion (Boulder, CO, USA; Chicago, IL, USA; Louisville, CO, USA; Madison, WI, USA; Oxford, England, UK)}

\author{Bettina Heim}
\affiliation{NVIDIA, Santa Clara, CA, USA}

\author{Garrett T. Hickman}
\affiliation{Infleqtion (Boulder, CO, USA; Chicago, IL, USA; Louisville, CO, USA; Madison, WI, USA; Oxford, England, UK)}

\author{Eric B. Jones}
\affiliation{Infleqtion (Boulder, CO, USA; Chicago, IL, USA; Louisville, CO, USA; Madison, WI, USA; Oxford, England, UK)}

\author{Ryan A. Jones}
\affiliation{Infleqtion (Boulder, CO, USA; Chicago, IL, USA; Louisville, CO, USA; Madison, WI, USA; Oxford, England, UK)}

\author{Pradnya Khalate}
\affiliation{NVIDIA, Santa Clara, CA, USA}

\author{Jin-Sung Kim}
\affiliation{NVIDIA, Santa Clara, CA, USA}

\author{Kevin W. Kuper}
\affiliation{Infleqtion (Boulder, CO, USA; Chicago, IL, USA; Louisville, CO, USA; Madison, WI, USA; Oxford, England, UK)}

\author{Martin T. Lichtman}
\affiliation{Infleqtion (Boulder, CO, USA; Chicago, IL, USA; Louisville, CO, USA; Madison, WI, USA; Oxford, England, UK)}

\author{Stephanie Lee}
\affiliation{Infleqtion (Boulder, CO, USA; Chicago, IL, USA; Louisville, CO, USA; Madison, WI, USA; Oxford, England, UK)}

\author{David Mason}
\affiliation{Infleqtion (Boulder, CO, USA; Chicago, IL, USA; Louisville, CO, USA; Madison, WI, USA; Oxford, England, UK)}

\author{Nathan A. Neff-Mallon}
\affiliation{Infleqtion (Boulder, CO, USA; Chicago, IL, USA; Louisville, CO, USA; Madison, WI, USA; Oxford, England, UK)}

\author{Thomas W. Noel}
\affiliation{Infleqtion (Boulder, CO, USA; Chicago, IL, USA; Louisville, CO, USA; Madison, WI, USA; Oxford, England, UK)}

\author{Victory Omole}
\affiliation{Infleqtion (Boulder, CO, USA; Chicago, IL, USA; Louisville, CO, USA; Madison, WI, USA; Oxford, England, UK)}

\author{Alexander G. Radnaev}
\affiliation{Infleqtion (Boulder, CO, USA; Chicago, IL, USA; Louisville, CO, USA; Madison, WI, USA; Oxford, England, UK)}

\author{Rich Rines}
\affiliation{Infleqtion (Boulder, CO, USA; Chicago, IL, USA; Louisville, CO, USA; Madison, WI, USA; Oxford, England, UK)}

\author{Mark Saffman}
\affiliation{Infleqtion (Boulder, CO, USA; Chicago, IL, USA; Louisville, CO, USA; Madison, WI, USA; Oxford, England, UK)}
\affiliation{Department of Physics, University of Wisconsin-Madison, Madison, WI, USA}

\author{Efrat Shabtai}
\affiliation{NVIDIA, Santa Clara, CA, USA}

\author{Mariesa H. Teo}
\affiliation{Pritzker School of Molecular Engineering, University of Chicago, Chicago, IL, USA}

\author{Bharath Thotakura}
\affiliation{Infleqtion (Boulder, CO, USA; Chicago, IL, USA; Louisville, CO, USA; Madison, WI, USA; Oxford, England, UK)}

\author{Teague Tomesh}
\affiliation{Infleqtion (Boulder, CO, USA; Chicago, IL, USA; Louisville, CO, USA; Madison, WI, USA; Oxford, England, UK)}

\author{Angela K. Tucker}
\affiliation{Infleqtion (Boulder, CO, USA; Chicago, IL, USA; Louisville, CO, USA; Madison, WI, USA; Oxford, England, UK)}

\date{\today}

\begin{abstract}
We report on the fault-tolerant operation of logical qubits on a neutral atom quantum computer, with logical performance surpassing physical performance for multiple circuits including Bell states (12x error reduction), random circuits (15x), and a prototype Anderson Impurity Model ground state solver for materials science applications (up to 6x, non-fault-tolerantly). The logical qubits are implemented via the [[4, 2, 2]] code (C\textsubscript{4}). Our work constitutes the first complete realization of the benchmarking protocol proposed by \textit{Gottesman 2016} \cite{gottesman2016quantum} demonstrating results consistent with fault-tolerance. In light of recent advances on applying concatenated C\textsubscript{4}/C\textsubscript{6} detection codes to achieve error correction with high code rates and thresholds, our work can be regarded as a building block towards a practical scheme for fault tolerant quantum computation. Our demonstration of a materials science application with logical qubits particularly demonstrates the immediate value of these techniques on current experiments.
\end{abstract}

\maketitle

\section{\label{sec:level1}Introduction}

\begin{figure}[!ht]
\centering
\includegraphics[width=0.5\textwidth]{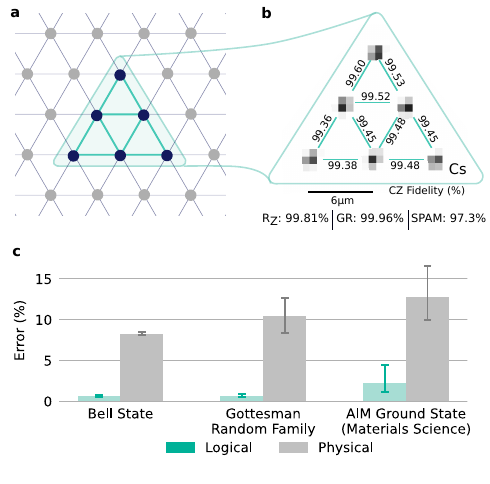}
\caption{\textbf{Fault-tolerant operation of logical qubits encoded with the [[4, 2, 2]] error detection code}. (a) This work utilizes 6 qubits in a hexagonal lattice. (b) Image of Cesium atoms arranged in a hexagonal lattice, with $6~\mu\rm m$ qubit spacing. CZ gate fidelities, post-selected for atom-loss, are shown for all connections.  Array-median fidelities for $R_Z$, $GR$ and SPAM are also presented. (c) Across multiple experiments, we observe significantly lower error with our encoded logical qubits than our unencoded physical qubits. The error metrics for the Bell state, Gottesman Random Family, and AIM Ground State Materials Science application are (respectively) state infidelity, total variation distance, and relative error w.r.t. the ground state.  
\label{fig:Fig1}
}
\end{figure}

As quantum computing platforms advance in qubit count and gate fidelity, they are approaching the requirements for scalable fault-tolerant operation with the capability to solve classically-intractable problems. Among these quantum computing platforms, neutral atoms have emerged as a leading qubit modality following demonstrations of algorithms \cite{graham2022multi} and  logical qubits \cite{bluvstein2024logical, reichardt2024logical}. In this paper, we demonstrate fault-tolerant operation of logical qubits on Infleqtion's quantum computer, which features individually optically-addressable Cesium atoms as qubits. Our experiments build atop progress on neutral atom hardware~\cite{radnaev2024universal}. 

In this paper, we present experimental results on our neutral atom quantum computer for the following:
\begin{itemize}
    \item Sec.~\ref{sec:gottesman}: The first fault-tolerant realization of the full Gottesman benchmarking protocol \cite{gottesman2016quantum} on two logical qubits. For the family of random circuits tested, we find that logical encoded circuits have a 15x average reduction in Total Variation Distance (TVD) over the physical unencoded circuits (10.5\% vs. 0.7\%) for $\ket{00}$ input state.
    \item Sec.~\ref{sec:matsci_app}: Demonstration of a prototype materials science application, with the [[4, 2, 2]] code. Specifically, we perform ground state preparation for the single-impurity Anderson Impurity Model (AIM) using the Hamiltonian Variational Ansatz \cite{wecker2015progress}. While our realization is not fault-tolerant, the logical encoded circuits achieve $6\times$ reduction in relative error over the physical unencoded circuits (2.3\% vs. 13\%).
    \item Sec.~\ref{sec:bell_ghz}: Specific benchmarking of the logical fidelity of a fault-tolerant realization of a Bell state. By performing state tomography, we extract a physical fidelity of 91.7(2)\%, and calculate logical fidelities of 96.5(3)\% and 99.3(1)\% when post-selecting on one or both stabilizers. We thus infer the logical encodings reduce error by up to 12.4(23)x.
\end{itemize}

Aggregated summary of these results are presented in Fig. \ref{fig:Fig1}. 
Noise-realistic simulations of these experiments are available as Jupyter notebooks leveraging Superstaq's compilation capabilities \cite{campbell2023superstaq} alongside NVIDIA's CUDA-Q \cite{cuda-q} library providing GPU acceleration.

We begin in Sec.~\ref{sec:prior_work} with a summary of prior experiments with the [[4, 2, 2]] code, followed in Sec.~\ref{sec:ft_operation} by an overview of the fault-tolerant gateset available for this code.

\section{Prior Work}  \label{sec:prior_work}

In response to experimental progress towards scalable quantum computing, \textit{Gottesman 2016} \cite{gottesman2016quantum} proposed a set of experiments for demonstrating operation of $k=2$ logical qubits using the [[$n=4$, $k=2$, $d=2$]] code \cite{grassl1997codes, vaidman1996error}, which uses 5 ($n=4$ data + 1 ancilla) physical qubits to achieve distance $d = 2$, corresponding to error detection. Importantly, the proposal in \textit{Gottesman 2016} is fault-tolerant, meaning that any single-qubit error is either detectable or inconsequential (has no logical effect). While the proposed experiments would demonstrate many of the key characteristics towards future fault-tolerant quantum computers, they lack certain key ingredients---notably error correction ($d \ge 3$), repeated syndrome extraction via mid-circuit measurement \& reset, and a universal gateset.

Nonetheless, the relatively modest requirements of \textit{Gottesman 2016} have invited several experimental demonstrations of the [[4, 2, 2]] code, which are summarized  in Tab.~\ref{tab:prior_res_422}. These previous experiments span superconducting (IBM), trapped ion (UMD, Quantinuum), photonic (USTC), and---during the preparation of this manuscript---neutral atom (Atom Computing) qubits. While many of these experiments have followed the approach suggested by \textit{Gottesman 2016}, most have been unable to fault-tolerantly prepare the $\ket{00}_L$ logical state, which requires 5 physical qubits in ring or star connectivity, making use of a flag qubit. To this end, our work relies on a hexagonal lattice atom arrangement, which conveniently enables a 5-qubit ring for $\ket{00}_L$ preparation, as well as the necessary connectivity for the materials science application in Sec.~\ref{sec:matsci_app}.

As expanded upon in Sec.~\ref{sec:gottesman}, our work demonstrates the first demonstration of the full benchmarking protocol proposed by \textit{Gottesman 2016} \cite{gottesman2016quantum} where every logical qubit circuit performs at least as well as the physical qubit circuits. However, we emphasize that  scalable fault-tolerant quantum computation requires three key characteristics that are not realized in this work. First, while our work measures the [[4, 2, 2]] code's $ZZZZ$ stabilizer (implicitly, through parities of bitstrings in the computational basis), we do not measure the code's other stabilizer, $XXXX$. As such, our experiment does not suppress errors that trigger the $XXXX$ stabilizer---though this is possible in principle due to the fault-tolerant constructions. Second, our experiment does not use mid-circuit measurement (MCM) and reset, which would be necessary for repeated syndrome extraction. This is likely a strict requirement for utility-scale quantum computation. Third, our demonstration does not span a universal gate set. We note that recent work by IBM \cite{gupta2024encoding} has achieved a CZ magic state for the [[4, 2, 2]] code, with better-than-breakeven fidelity.

[[4,2,2]] code is also known as the $C_4$ code. While it is an error detection code, concatenation of such codes leads to error correction codes \cite{knill2005quantum, gottesman2016quantum} such as the C4/C6 code. Recent work has demonstrated that such concatenated structures admit fault-tolerant quantum computing schemes with high thresholds and 1-2 orders of magnitude higher code rate than baseline surface code approaches \cite{yoshida2024concatenate, goto2024many}.

On its own, the [[4,2,2]] code only detects errors, however, this capability proves valuable for executing quantum algorithms at the hardware's operational limits. This aligns with recent demonstrations where quantum error detection protocols significantly enhanced the fidelity of QAOA implementations \cite{he2024performance}. The complementary deployment of error correction and detection strategies creates a pathway for pushing quantum computers to their operational boundaries while maintaining computational integrity. Such robust error management will be crucial for extending the prototype quantum chemistry application demonstrated here toward solving real-world problems such as \textit{in silico} protein modeling with applications to therapeutic drug development \cite{ramesh2024quantum}.

\begin{table*}[htp!]
    \begin{tabular}{|p{0.7 in}|p{0.85in}|p{0.5 in}|p{0.68 in}|p{0.35 in}|p{3.1 in}|} \hline 
          Date &  Platform &  FT $\ket{00}_L$  & XXXX Stabilized & MCM & Applications of [[4,2,2]]\\ \hline 
          Dec 2024 [This Work] & Neutral Atom (Infleqtion) & Yes & No & No &  Realization of Gottesman benchmarking protocol, Bell state tomography, material science application\\ \hline 
          Nov 2024 \cite{reichardt2024logical}& Neutral Atom (Atom Comp.) & Yes & Yes & No & Demonstration of entanglement between 24 logical qubits; repeated error detection and loss correction for circuits with repeated CZ and dual CZ gates, and random circuits with up to 3 rounds of error detection\\ \hline 

          Sep 2024 \cite{van2024end}& Trapped Ion (Quantinuum) & No& Yes& Yes& Implementation of ground state preparation circuit in the end-to-end execution of a quantum chemistry algorithm\\ \hline 
          Jan 2024 \cite{gupta2024encoding}& SC (IBM)& N/A & N/A & N/A & Focused on preparing magic state with [[4, 2, 2]] code; demonstrates better logical than physical performance\\ \hline 
          Jul 2022 \cite{sun2022optical}&  Photons (USTC) & No& No& No& Demonstration of pseudothreshold for a complete circuit\\ \hline 
          Jun 2022 \cite{zhang2022comparative}&  SC (IBM) & No & No & No & Implementation of VQE circuit for finding the ground state energy of a Hydrogen molecule\\ \hline 
          Dec 2021 \cite{cane2021experimental}& SC (IBM) & No & No & No & Gottesman-inspired benchmarking protocol (certain circuits fell short of the fault tolerance criterion)\\ \hline 
          Aug 2020 \cite{urbanek2020error}& SC (IBM) & Yes & No & No & Implementation of VQE circuit for finding the ground state energy of a Hydrogen molecule\\ \hline 
          Aug 2020 \cite{kole2020resource}& SC (IBM) & No & No & No & Gottesman-inspired benchmarking protocol with an expanded gateset (certain circuits fell short of the fault tolerance criterion)\\ \hline 
          Feb 2019 \cite{harper2019fault}& SC (IBM) & No & No & No & Gottesman-inspired benchmarking protocol  characterizing logical gateset, but not SPAM\\ \hline 
          Nov 2018 \cite{willsch2018testing}& SC (IBM) & No & No & No & Gottesman benchmarking protocol with $T=10$, $r=6$, $p=3$, excluding $|00\rangle_L$ state preparation\\ \hline 
          Sep 2018 \cite{vuillot2017error}& SC (IBM) & Yes & No & No & Gottesman-inspired benchmarking protocol of various state preparation circuits (some circuits fell short of the fault tolerance criterion)\\ \hline
          Jun 2018 \cite{roffe2018protecting}& SC (IBM) & No & No & No & Implementation of [[4,2,2]] code as a coherent parity check code\\ \hline
          Oct 2017 \cite{takita2017experimental}& SC (IBM) & No & No & No & Demonstration of improved state preparation circuits using the [[4,1,2]] subsystem code (only one of the two LQs is fault tolerant)\\ \hline
          Oct 2017 \cite{linke2017fault} & Trapped Ions (UMD) & No & Yes (Separately) & No & Demonstration of improved state preparation and single qubit gates using the [[4,1,2]] subsystem code (only one of the two LQs is fault tolerant)\\\hline
    \end{tabular}
    \caption{Previous experimental realizations of the [[4,2,2]] code.}
    \label{tab:prior_res_422}
\end{table*}

\section{Fault-Tolerant Operation} \label{sec:ft_operation}

\begin{figure*}
    \centering
    \includegraphics[width=0.99\linewidth]{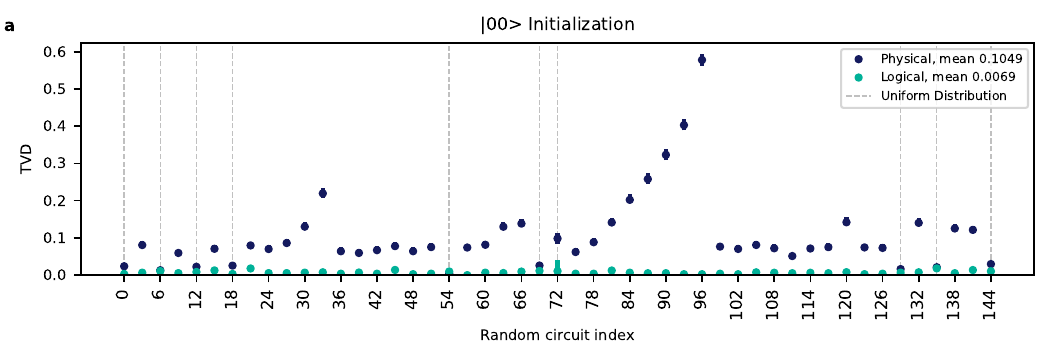}
    \caption{Gottesman benchmarking for $\ket{00}$ fault-tolerant logical initialization. Points indicate raw measured TVDs.  Error bars indicate the narrowest 68\% confidence intervals, based on Monte Carlo sampling of a Dirichlet distribution based on the observed counts. We find a 15x reduction in error from a TVD of 10.5\% for physical qubits to 0.7\% for logical qubits. Each index was initially run with approximately 1,050 shots in both the physical and logical qubit configurations. An additional 7,000 shots were run on the indices where physical and logical TVD envelopes were intersecting.}
    \label{fig:00_init}
\end{figure*}

The [[4, 2, 2]] code uses 4 physical qubits to encode 2 logical qubits with a distance of $d=2$, meaning that a single error can be detected (but not corrected). The code is a Calderbank-Shor-Steane (CSS) stabilizer code with stabilizers $XXXX$ and $ZZZZ$. This gives rise to the logical codewords:
\begin{eqnarray}
    \ket{00}_L = \left(\ket{0000} + \ket{1111}\right)/\sqrt{2} \nonumber \\
    \ket{01}_L = \left(\ket{0011} + \ket{1100}\right)/\sqrt{2} \nonumber \\
    \ket{10}_L = \left(\ket{0101} + \ket{1010}\right)/\sqrt{2} \nonumber \\
    \ket{11}_L = \left(\ket{0110} + \ket{1001}\right)/\sqrt{2}
\end{eqnarray}

This encoding can also be interpreted in reverse as a decoding. For instance, measurement of $\ket{0101}$ would indicate the $\ket{10}_L$ logical state. Measurement of a bitstring with odd parity is outside of the logical codespace and indicates a detected error.

Although only 4 physical qubits encode the underlying logical data, fault tolerant operation is possible with a 5th  physical qubit in ring topology (specifically, acting as a flag qubit for $\ket{00}_L$ initialization) or star topology \cite{urbanek2020error}. Tab.~\ref{tab:ft_operations} in the Appendix shows explicit fault-tolerant circuits for state preparation, logical gate, and measurement operations. For each operation, fault tolerance can be verified by considering a physical bit-flip and/or phase-flip error at any location in the encoded circuits and confirming that the error cannot propagate.

\textit{State preparation.} All three fault tolerant state preparation circuits are shown in Tab.~\ref{tab:ft_operations}. The \texttt{PREP\_00} circuit includes a flag qubit necessary for fault-tolerance to verify that $q_0 + q_3$ has even parity; odd parity indicates an error has been detected. The \texttt{PREP\_0+} and \texttt{PREP\_BELL} physical circuits correspond to preparation of two physical Bell states on opposite pairs of a square.

\textit{Logical Gates.} All of the gates we consider have transversal implementations, as depicted. The $Y$ gate on either qubit can be formed as the concatenation of the $X$ and $Z$ gate sequences on that qubit. The SWAPs in the logical \texttt{HH} and \texttt{CX} circuits are implemented virtually (i.e. relabeling) rather than via physical gates; thus, no physical two-qubit gates are required after the initial state preparation.

\textit{Measurement.} In the encoded circuits, in addition to the physical measurement, we perform postselection to discard shots with odd parity across the four measured qubits, since odd parity bitstrings are outside the logical codespace as described previously. Recall that there is also post-selection on the flag qubit parity measurement in the \texttt{PREP\_00} circuit. This could in principle be performed under a repeat-until-success scheme with feedforward mid-circuit measurement and reset. In our experiments, we perform all measurements terminally.

\subsection{Compilation to Neutral Atom Gateset}
We used Superstaq \cite{campbell2023superstaq} integrated with NVIDIA CUDA-Q \cite{cuda-q} to compile each encoded circuit to the neutral atom gateset which include $GR_{\theta, \phi}$ global rotation gates that rotate every qubit by $\theta$ radians along the $\phi$ axis in the XY plane, single-site $R_z$ rotation gates (e.g. $Z$, $S$, and $S^{\dagger}$), and CZ entangling gates between connected qubits. The global rotation gate is a distinct aspect of the neutral atom platform. Since it acts even on qubits that would otherwise be idling, the compiler must apply appropriately placed sandwiches of $GR$ and $R_z$ rotations such that the intended gate is applied in net to each qubit.

This global interaction behavior means that our logical $X$, $Z$, and \texttt{HH} compiled circuits act on all five physical qubits, including the ancilla parity-check flag qubit used by \texttt{PREP\_00}, even though there are only four data qubits. This is necessary to ensure that the flag qubit is not disturbed by the logical gates. If our logical state is initialized to $\ket{0+}$ or the $(\ket{00} + \ket{11}) / \sqrt{2}$ Bell state, the flag qubit is not needed, allowing for a simpler compilation of the logical \texttt{HH} gate (also shown in Tab.~\ref{tab:ft_operations}).

\section{Gottesman Protocol Verification} \label{sec:gottesman}

To verify the performance and fault-tolerant behavior of our logical qubits, we executed the full benchmarking protocol proposed by Gottesman \cite{gottesman2016quantum}. Under this protocol, we tested subfamilies of logical (encoded) vs. physical (unencoded) circuits, parametrized by $T=8$ (maximum depth of gate layers), $r=2$ (number of different circuits to draw for each kind), and $p=4$ (maximum periodicity). For each gate layer, we draw from the 8 different logical gates shown in Tab.~\ref{tab:ft_operations}. We exclude \texttt{II}, which is a no-op for both encoded and unencoded circuits.

Per the Gottesman protocol, we consider two types of circuits, described below. Each circuit of depth $t$ comprises $t+2$ fault-tolerant logical steps: state preparation, $t$ logical operations, and finally measurement. \textit{Type 1}: For each $t \in [1, ..., T]$ we sample $T$ random layers. There are $\leq r(T+1)$ Type 1 circuits chosen. \textit{Type 2}: For each period $q \in [1, ..., p]$, draw $q$ random gate layers and repeat them $\lfloor T/q \rfloor$ times. There are at most $rT(\ln{p} + 1)$ circuits chosen from Type 2. These circuits are designed to stress-test coherent errors that would accumulate from the periodic repetition.

Our $T=8, r=2, p=4$ selection resulted in 49 base circuits for benchmarking, listed in full in Appendix~\ref{app:gottesman_circuits}. For each base circuit, we tested with the fault-tolerant \texttt{PREP\_00}, \texttt{PREP\_0+}, and \texttt{PREP\_BELL} initializations, resulting in a total of 147 circuits tested. We ran each circuit with fault-tolerant encoded circuits (logical qubits) and unencoded circuits (physical qubits). The unencoded circuits were executed on the second and third physical qubits in our five-qubit ring; this pair generally had the best CZ fidelities, ensuring that our comparison to unencoded circuits is stringent.

Per the Gottesman protocol's prescription, our figure of merit is the Total Variation Distance (TVD) between the experimental results and the ideal distribution for each of the 147 circuits. We computed 68\% confidence intervals for these TVD metrics using a Markov Chain Monte Carlo approach described in Appendix~\ref{app:tvd_error}. We executed approximately 1,050 shots for each of the 147 benchmarking circuits, in both the logical encoded and physical unencoded configurations.

Our primary interest is in the performance for the \texttt{PREP\_00} circuits, both because is the most challenging state preparation (five CZs for encoded circuits, versus zero for unencoded) and because it aligns most closely with typical quantum computing workflows which begin with the all-zeros input state. After running the initial 1,050 shots per circuit, we found nine indices for \texttt{PREP\_00} initialization where the confidence intervals for logical vs. physical qubits were overlapping (some favorably and some unfavorably for the logical qubits). We utilized our remaining shot budget to perform 7,000 additional shots per circuit on these nine indices. Thus, in total, our Gottesman benchmarking experiments spanned over 450,000 shots.

The results for the 49 random circuits initialized to \texttt{PREP\_00} circuits are shown in Fig.~\ref{fig:00_init}. The logical qubits have a mean TVD of 0.7\% versus 10.5\% for the physical qubits, corresponding to a 15x error reduction via the fault-tolerant encoding. To conclude that the results are consistent with fault-tolerance, we must also examine each circuit index. However, we note that 10 of the 49 random circuits have an ideal output of the uniform distribution (25\% probability for each of the four bitstrings). These cases are pathological, because it is also the stationary state for depolarizing noise and the physical qubits could ``accidentally'' achieve low TVD for these circuits.

On the other hand, these cases also detect error mechanisms that would not be penalized in other circuits for the logical qubits. For example, if our $\ket{00}_L$ preparation produced $(\alpha \ket{0000} + \beta \ket{1111}) / \sqrt{2}$ with $\alpha \neq \beta$, only the circuits with uniform distribution would penalize this error. As such, it is still important to include these circuits in our benchmarking. Our device-realistic simulations, detailed in Appendix~\ref{app:simulation}, indicated that our logical qubits should achieve TVD lower than physical qubits for all 147 indices, but could require nearly 100,000 shots for some of the circuits to sufficiently resolve a separation. Given a limited shot budget, we therefore desire that:
\begin{itemize}
    \item the logical qubits outperform physical qubits (with a separation between TVD uncertainty envelopes) for all non-uniform-distribution circuits
    \item the logical qubits at least equal the performance of physical qubits (intersecting TVD uncertainty envelopes are acceptable) for all uniform-distribution circuits.
\end{itemize} 
Indeed, the results shown in Fig.~\ref{fig:00_init} achieve both objectives and we conclude that our results are consistent with fault-tolerance.

\begin{figure}
    \centering
    \includegraphics[width=0.99\linewidth]{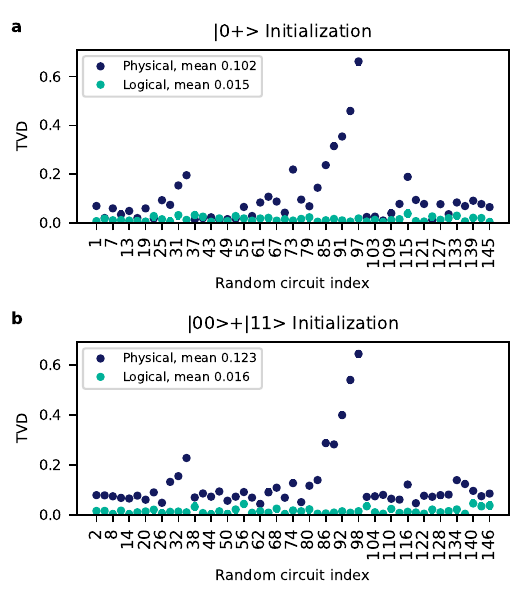}
    \caption{Gottesman benchmarking for fault-tolerant \textbf{a} $\ket{0+}$ and \textbf{b} $\ket{\phi^+} = (\ket{00} + \ket{11}) / \sqrt{2}$ Bell state  logical initialization. For the $\ket{0+}$ state, we observe a 6.7x reduction in mean TVD from physical (10.2\%) to logical (1.5\%) qubits. For the $\ket{\phi^+} = (\ket{00} + \ket{11}) / \sqrt{2}$ state, we observe a 7.7x reduction in mean TVD from physical (12.3\%) to logical (1.6\%) qubits.}
    \label{fig:other_gottesman_inits}
\end{figure}

We also include results for the 98 (49+49) circuits for \texttt{PREP\_0+}, and \texttt{PREP\_BELL} initializations in Fig.~\ref{fig:other_gottesman_inits} respectively. The experimental results indicate a 6.7x and 7.7x mean reduction in TVD via the logical qubit encodings. We note that the Bell state initialization is directly relevant to applications that require initially entangled resource states.

\section{Prototype Materials Science Application: Anderson Impurity Model} \label{sec:matsci_app}

Ground state quantum chemistry---computing total energies of molecular configurations to within chemical accuracy---is perhaps the most highly-touted industrial application of fault-tolerant quantum computers. However, the large gate counts required for precise quantum phase estimation, combined with the lack of a clear exponential speedup against classical methods and the temporal overhead for quantum error correction, may push out the timeline for industrially-useful quantum computational chemistry \cite{berry2024rapid, lee2023evaluating, babbush2021focus}. Meanwhile, it has been long known that simulating the dynamics of many-body quantum systems is the most natural application of quantum computers--- conferring an exponential speedup against classical methods--- and was the application for which they were originally envisioned \cite{feynman2018simulating}. State-of-the-art classical electronic structure methods for computing dynamical properties of materials, such as photoemission spectra and magnetic susceptibilities, are formulated in terms of Green's functions, which can be computed efficiently via quantum simulation on a quantum computer \cite{jones2024dynamic}. Specifically, dynamical mean-field theory (DMFT) allows to account for the effect of strong, localized correlations on the electronic structure in a manner that can be made self-consistent with high-fidelity quasiparticle theories like $GW$ (Green's function-screened Coloumb interaction) theory, which are scalable classically \cite{zhu2021ab}. DMFT+$GW$ frameworks have begun to be able to describe the physics of unconventional superconducting \cite{kurleto2023flat, acharya2022role} materials, and are limited in their accuracy by the exponential classical scaling of the DMFT impurity problem at low temperatures \cite{bravyi2017complexity}. Here we take a first step towards running DMFT on a fault-tolerant quantum computer by preparing the ground state of the minimal single-impurity Anderson model (SIAM) on two logical qubits encoded in the [[4, 2, 2]] code for a range of realistic parameters. In order to close the DMFT loop, one needs to further compute the one-particle Green's function, which we leave to future work.

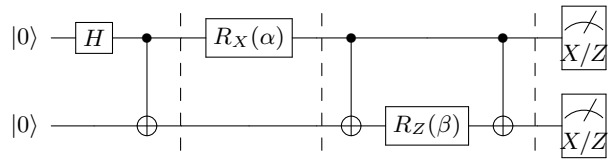
\begin{figure}[!t]
    \centering
$$
\Qcircuit @C=1em @R=1em {
& \lstick{\ket{0}} & \gate{H}  & \ctrl{1} & \qw \barrier[-2.8em]{1} & \gate{R_X(\alpha)} &\qw \barrier[-1.2em]{1} & \ctrl{1} & \qw & \ctrl{1} & \qw \barrier[-2em]{1} & \meterB{X/Z} \\
& \lstick{\ket{0}} & \qw & \targ & \qw & \qw & \qw & \targ    & \gate{R_Z(\beta)} & \targ & \qw & \meterB{X/Z}
}
$$
\caption{Physical (unencoded) ground state preparation circuit for Anderson Impurity Model, using a restricted Hamiltonian Variational Ansatz. Note that when measuring in the $Z$-basis, the $ZZ(\beta)$ component of the ansatz (implemented between the second and third barriers) has no logical effect and so is elided.}
    \label{fig:aim_logical}
\end{figure}

The SIAM Hamiltonian is
\begin{equation}\label{eq:siam}
\begin{split}
H_{\text{SIAM}} &= h\sum_{\sigma=\downarrow}^{\uparrow}c^{\dagger}_{I\sigma}c_{I\sigma} + U c^{\dagger}_{I\uparrow}c_{I\uparrow}c^{\dagger}_{I\downarrow}c_{I\downarrow} \\
&+ \epsilon \sum_{\sigma=\downarrow}^{\uparrow} c^{\dagger}_{B\sigma}c_{B\sigma} + V\sum_{\sigma=\downarrow}^{\uparrow}(c^{\dagger}_{I\sigma}c_{B\sigma} + h.c.),
\end{split}
\end{equation}
where $c^{\dagger}_{i\sigma} (c_{i\sigma})$ are operators that create (annihilate) an electron with spin component $\sigma$ on either the impurity ($i=I$) or bath ($i=B$) site. In general Eq.~\ref{eq:siam} requires four qubits to be simulable on a quantum computer. At half-filling, where the chemical potential is taken to be $h=-U/2$ and the bath energy is taken to be $\epsilon=0$ \cite{jaderberg2020minimum}, and where the SIAM retains enough predictive power to exhibit, among other phenomena, a metal insulator transition \cite{potthoff2001two}, one can map Eq.~\ref{eq:siam} to a Hamiltonian acting on two qubits. To show this, we begin by ordering the orbitals as $(I\uparrow, B\uparrow, I\downarrow, B\downarrow) \rightarrow (0, 1, 2, 3)$. Under the Bravyi-Kitaev mapping \cite{bravyi2002fermionic}, Eq.~\ref{eq:siam} becomes
\begin{equation}\label{eq:bk_map}
H_{\text{BK}} = \frac{U}{4}(Z_0Z_2 - 1) + \frac{V}{2}(X_0 -X_0Z_1 - Z_1X_2Z_3 + X_2).
\end{equation}
The four fermionic basis states that support ground states of Eq.~\ref{eq:siam} at half-filling are $\{\ket{0101}, \ket{0110}, \ket{1001}, \ket{1010}\}$, which under Bravyi-Kitaev map to $\{\ket{0111}, \ket{0110}, \ket{0011}, \ket{0010}\}$ and which have the form $\ket{0 z_2 1 z_0}$. Hence, Eq.~\ref{eq:bk_map} stabilizes qubits 1 and 3, reducing the problem to a two-qubit Hamiltonian defined by the Coulomb interaction $U$ and hybridization strength $V$
\begin{equation}\label{eq:2q_ham}
H_{(U, V)} = U (Z_0 Z_2 - 1) / 4 + V (X_0 + X_2),
\end{equation}
which acts on the basis set $\ket{z_2z_0} = \{\ket{00}, \ket{01}, \ket{10}, \ket{11}\}$. While similar qubit reduction techniques have been used in the context of the $H_2$ molecule \cite{o2016scalable}\cite{urbanek2020error}, the reduction herein for the SIAM at half-filling is new. Each term in Eq.~\ref{eq:2q_ham} can be measured via the [[4, 2, 2]] code since both $Z$ and $X$ basis measurements are available in the logical gateset. We consider a restricted form of the Hamiltonian Variational Ansatz \cite{wecker2015progress},
\begin{equation}
\mathcal{U}(\alpha, \beta) = e^{-i \beta Z_0 Z_2/2} e^{-i \alpha X_0 / 2}
\end{equation}
acting on the $\ket{\phi^+}$ Bell state. The corresponding physical (i.e., unencoded) circuit for ground state preparation is shown in Fig.~\ref{fig:aim_logical}, which we confirmed was sufficiently expressible and trainable in its ability to find the true ground state by training it classically.

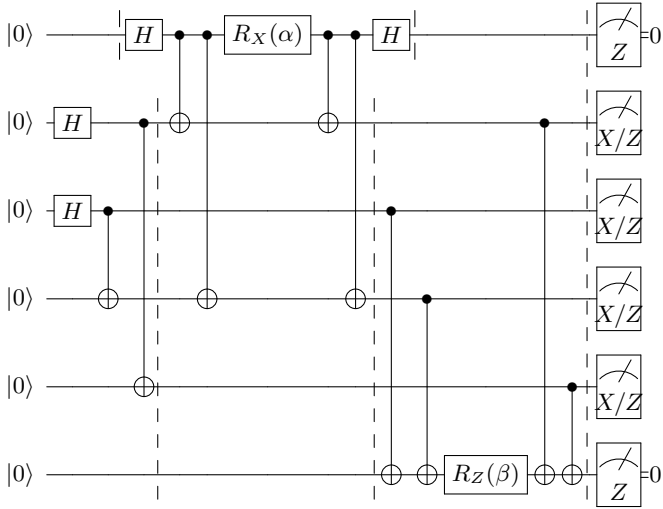
\begin{figure}[!h]
    \centering
$$
\Qcircuit @C=0.3em @R=1em {
& \lstick{\ket{0}} & \qw & \qw  \barrier[-1em]{0} & \gate{H} & \ctrl{1} & \ctrl{3} & \gate{R_X(\alpha)} & \ctrl{1} & \ctrl{3} & \gate{H} \barrier[-0.5em]{0} & \qw & \qw & \qw & \qw & \qw \barrier[-1.3em]{5} & \meterB{Z} & \cw & 0\\
& \lstick{\ket{0}} & \gate{H} & \qw & \ctrl{3} \barrier[-0.9em]{4} & \targ & \qw & \qw & \targ & \qw \barrier[-0.7em]{4} & \qw & \qw & \qw & \ctrl{4} & \qw & \qw &\meterB{X/Z}\\
& \lstick{\ket{0}} & \gate{H} & \ctrl{1} & \qw & \qw &\qw & \qw & \qw & \qw & \ctrl{3} & \qw & \qw & \qw & \qw & \qw & \meterB{X/Z}\\
& \lstick{\ket{0}} & \qw & \targ & \qw & \qw & \targ & \qw & \qw & \targ & \qw & \ctrl{2} & \qw & \qw & \qw & \qw & \meterB{X/Z}\\
& \lstick{\ket{0}} & \qw & \qw & \targ & \qw & \qw & \qw & \qw & \qw & \qw & \qw & \qw & \qw & \ctrl{1} & \qw & \meterB{X/Z}\\
& \lstick{\ket{0}} & \qw & \qw & \qw & \qw & \qw & \qw & \qw & \qw & \targ & \targ & \gate{R_Z(\beta)} & \targ & \targ & \qw & \meterB{Z} & \cw & 0\\}
$$
\caption{Logical (encoded) ground state preparation circuit for Anderson Impurity Model, using a restricted Hamiltonian Variational Ansatz. Though not fully fault tolerant, the use of ancillary flag qubits enables us to detect if a single-qubit error occurs nearly anywhere in the circuit.}
    \label{fig:aim_encoded}
\end{figure}

The [[4, 2, 2]] logical encoding of this ansatz is shown in Fig.~\ref{fig:aim_encoded}. While the initial Bell state preparation is fault tolerant, the subsequent logical partial rotations $R_X(\alpha)$ and $ZZ(\beta)$ do not admit transversal implementations and so are not implemented fault-tolerantly. However, we find that by mediating both operations via ancillary ``flag'' qubits, we can still detect nearly all single-qubit errors which would otherwise result in logical errors in the encoded circuit outcome.

The subcircuit on the left-hand side of the first broken barrier in Fig.~\ref{fig:aim_encoded} encodes initial Bell state preparation using the fault-tolerant construction described in \cite{gottesman2016quantum}. The next circuit section (between the two broken barriers) implements the logical $R_X(\alpha)$ operation mediated through the top ancilla qubit, which is then (assuming no error occurs) returned to the $\ket0$ state. Similarly, the third circuit section (between the second broken barrier and the third barrier) implements the logical $ZZ(\beta)$ operation using the bottom ancilla. In this case we also incorporate a measurement of the $ZZZZ$ stabilizer, implemented by simply moving the controls of the final two CNOTs to the remaining two qubits (which has the same logical effect as flipping the ancilla according to the parity of all four data qubits). The final circuit section measures the ancillas in the $Z-$basis and the four primary qubits in either the $Z-$ and $X-$basis for subsequent post-selection and decoding. When measuring in the $Z$-basis, the $ZZ(\beta)$ component of the ansatz has no logical effect and the incorporated $ZZZZ$ stabilizer measurement is redundant, and so the prior circuit section and bottom ancilla qubit are elided.

Post-selecting based on the top ancilla's 0 state allows us to flag and throw away results in which certain phase errors occur during the logical $R_X(\alpha)$ sequence. Post-selecting based on the bottom ancilla's 0 state allows us to flag and throw away results in which a single-qubit $X$ error has occurred on it or one of the four primary data qubits. Though still not completely fault-tolerant, one can check that the only remaining single-qubit errors which can lead to logical errors in the circuit outcome are (1) bit flips in the immediate vicinity of the $R_X(\alpha)$ gate on the top ancilla qubit, and (2) phase flips in the immediate vicinity of the $R_Z(\beta)$ gate on the bottom qubit (both being indistinguishable from intentional shifts of either $\alpha$ or $\beta$). A single bit- or phase-flip error placed anywhere else in the circuit will either be detectable in the code or ancilla qubits, or otherwise have no effect on the decoded measurement distribution.

To assess the degree to which our hardware performance is under a threshold for improved state preparation by [[4, 2, 2]] error detection, we begin by training the circuits in Figs.~\ref{fig:aim_logical} and \ref{fig:aim_encoded} classically, using the energy expectation value,
\begin{equation}\label{eq:en_exp_val}
E_{(U, V)}(\alpha, \beta) = \bra{\phi^+} \mathcal{U}^{\dagger}(\alpha, \beta) H_{(U, V)} \mathcal{U}(\alpha, \beta) \ket{\phi^+},
\end{equation}
as the cost function. We trained the ansatz to minimize Eq.~\ref{eq:en_exp_val} in the absence of noise. To ensure that the ansatz was generally trainable and expressive, we ran optimizations for a two-dimensional grid of Hamiltonian parameter values, $U\in \{1, 5, 9\}$ and $V\in \{-9, -1, 7\}$, in units of electronvolts (eV). These parameter values are typical of the energy scales in which a DMFT loop would be performed \cite{potthoff2001two}. All optimizations converged to the exact ground state energy within a few dozen iterations at most. We justify classical training on the basis that (1) the logical circuits are small enough to train classically, and (2) for large systems there are more scalable state preparation heuristics that circumvent quantum variational loops, like matrix product state circuits \cite{berry2024rapid}.

\begin{figure}[!t]
\centering
\includegraphics[width=\columnwidth]{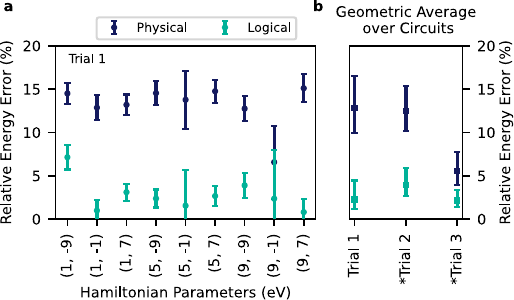}
\caption{Error detection in ground state preparation. \textbf{a} Relative error in hardware-sampled ground state energy versus exact ground state energy as a function of Hamiltonian parameters. Error bars represent one standard error of the mean. \textbf{b} Parameter-averaged relative energy error for three different data acquisition trials (see main text). Asterisks denote CUDA-Q trials. Averages and error bars are computed via the geometric mean and geometric standard deviation, respectively.
\label{fig:aim_energies}
}
\end{figure}

Once optimal angles, $(\alpha^*, \beta^*)_{(U, V)}$,  were found for each parameter set, $(U, V)$, we ran the corresponding circuits (Figs.~\ref{fig:aim_logical} and \ref{fig:aim_encoded}) on our quantum computer to compute the ground state energy estimate $
E^{\text{phys./log.}}_{(U, V)}(\alpha^*, \beta^*)$ in the presence of processor noise. To ensure the robustness of our results, we performed three rounds of data acquisition using slightly different workflows. In our first trial (Fig.~\ref{fig:aim_energies}a), we handled the variational training loop using the \textit{Qulacs} simulation package to classically emulate the unencoded circuits and the \textit{SciPy} implementation of the Broyden–Fletcher–Goldfarb–Shanno (BFGS) algorithm to optimize the angles to within machine precision \cite{suzuki2021qulacs}\cite{virtanen2020scipy}. We then chose a random embedding of the encoded circuits (Fig.~\ref{fig:aim_encoded}) into the processor topology in Fig.~\ref{fig:Fig1}. Similarly, we chose a random qubit pair within Fig.~\ref{fig:Fig1}, with characteristic error profile (Fig.~\ref{fig:Fig1}b), on which to execute the unencoded circuits (Fig.~\ref{fig:aim_logical}). All circuits for Trial 1 were compiled and optimized using Superstaq and submitted through Infleqtion's internal circuit submission system with 1,000 shots \cite{campbell2023superstaq}. The results are shown in Fig.~\ref{fig:aim_energies}. The exact ground state energies over the parameter set range over nearly an order of magnitude, from $\sim-2.27$ eV at $(1, -1)$ to $\sim-20.39$ eV at (9, -9). Hence, we plot the error in the energy expectation values relative to true ground state energies for all parameters. Despite this wide range of true energies, we find that on average (Fig.~\ref{fig:aim_energies}b), preparing the ground state with [[4, 2, 2]] error detection and flagging results in a 5.7x $\sim$ 6x reduction in relative ground state energy error versus the unencoded preparation. Moreover, the encoded circuits result in a reduction in relative error (lower energy) for every parameter set considered. This result represents the first time that a core subroutine of a strongly-correlated materials science application has been demonstrated to be improved via an error detection code. 

Trial 2 (Fig.~\ref{fig:aim_energies}b) also mapped the encoded and unencoded circuits to the hardware randomly, but leveraged integration with NVIDIA's CUDA-Q platform in order to generate ansatz circuits and perform the classical VQE optimization using GPU simulation. CUDA-Q circuits were then submitted through Superstaq, for compilation and optimization, to the quantum processor. In-spite of the circuit generation and submission changes involved in this integration, in addition to potential calibration drift, the error reduction outcomes in Trial 2 closely match those of Trial 1 (Fig.~\ref{fig:aim_energies}b).

In Trial 3, we attempted to stress test our processor's error correction capability by adding an additional ``hardware-aware'' compiler pass that chose the most highly-performant qubit pair, as measured by CZ fidelity, on which to run the unencoded circuits. Despite the resultant improvement in the unencoded relative error (Fig.~\ref{fig:aim_energies}b), the encoded relative error remains lower by a factor of about 2.6x. We conclude that our processor has an error profile that systematically benefits from the use of error detection in preparing accurate ground states of the SIAM. Moreover, the average relative error by which the encoded circuits deviated from the true ground state energy was less than 5\% in all trials, indicating a robust error detection protocol.

\section{2 Logical Qubits Bell State} \label{sec:bell_ghz}

\begin{figure}[h!]
\centering
\includegraphics[width=0.5\textwidth]{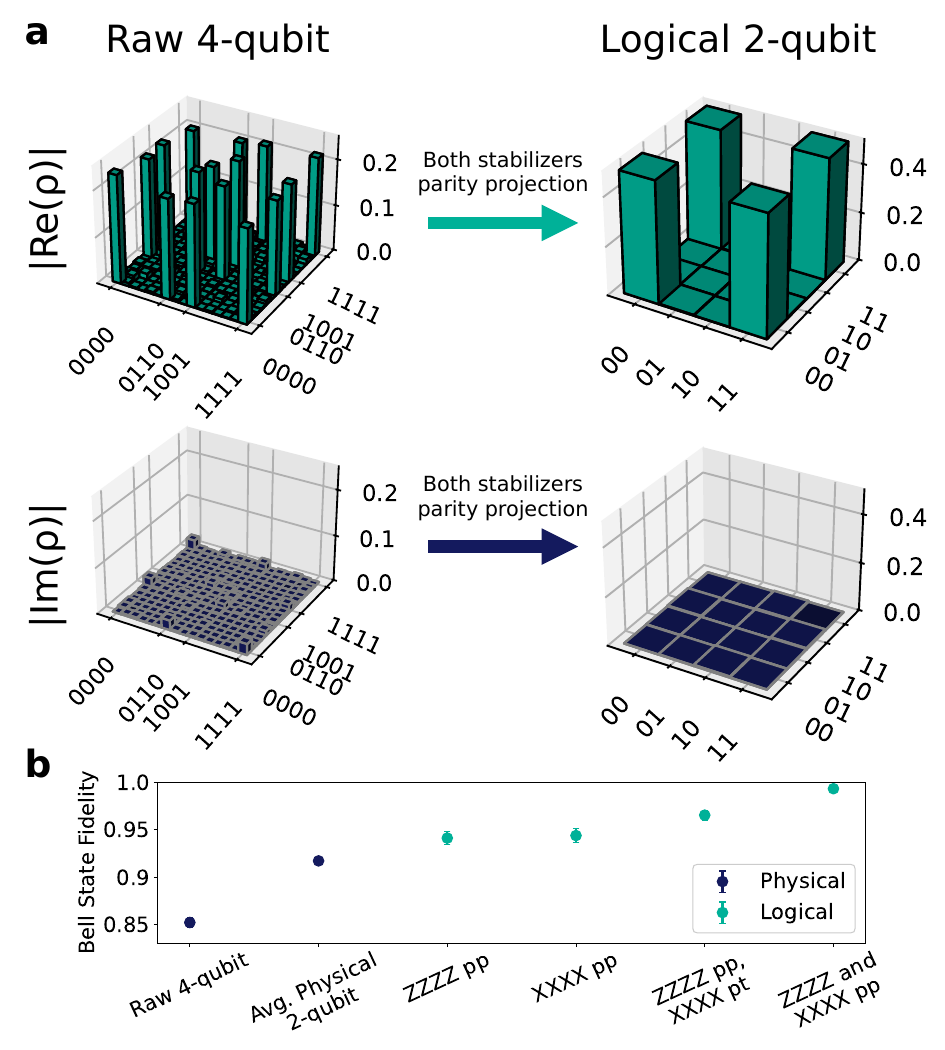}
\caption{Logical Bell state preparation. (a) Reconstructed density matrices from a tomographically-complete set of 81 measurements (each taken with 2000 shots prior to atom-loss post-selection) of the 4-qubit logical Bell state.  At left is the raw 4-qubit density matrix.  At right is the resulting density matrix after projecting the raw density matrix into the even-parity subspaces of both the ZZZZ and XXXX stabilizers. (b) Bell state fidelities for physical density matrices and logical density matrices with parity-projection (pp) and/or parity-trace (pt) operations (see main text).  Error bars indicate 95\% confidence intervals.
\label{fig:2LQ_Bell_Result}
}
\end{figure}

The previous sections have presented post-selected logical results by detecting errors with the ZZZZ stabilizer.  To get a deeper understanding of the effects of post-selection on our logical states, we characterize the logical Bell state $\ket{\phi^+} = (\ket{00} + \ket{11}) / \sqrt{2}$ prepared using the fault-tolerant gate set by performing state tomography on the underlying physical qubits.  To do so, we perform a tomographically-complete set of 81 measurements which takes the qubits out of the logical code space, but provides full information about the physical 4-qubit density matrix.

To both reconstruct the density matrix and obtain an uncertainty on the state fidelity, we utilize a reconstruction protocol \cite{faist2016tomography} that uses the Metropolis-Hastings algorithm to do a random walk among likely density matrices that match the experimental results.  Since the circuit that produces the logical Bell state also produces 2 physical Bell states, we can extract both logical and physical Bell fidelities from the most-likely density matrix output of the Metropolis-Hastings search.  The results are shown in Fig.~\ref{fig:2LQ_Bell_Result}.
Physical and Logical Bell fidelities can be found by projecting into or partially tracing over subspaces within the 4-qubit state space.  For instance, a physical pair's Bell fidelity may be obtained by performing partial traces over the other two qubits.  Doing so yields an average physical fidelity of 91.7(2)\%.

From the reconstructed density matrix, we can also now recreate the ZZZZ post-selection implemented in the previous sections.  Projecting our density matrix into the even-parity subspace of the ZZZZ stabilizer alone yields a Bell fidelity of 94.1(3)\%, already exceeding the physical Bell fidelity.  However, since our Z-basis measurements do not yield XXXX parity information, we effectively trace over this parity subspace. Doing so to the reconstructed density matrix yields a logical fidelity of 96.5(3)\%.  Finally, we calculate the Bell fidelity in the logical subspace that future experiments with access to both stabilizers may achieve after stabilization to be 99.3(1)\%, which is a 12.4(23)x improvement over the physical Bell state.  We note that this doubly-stabilized fidelity may currently be limited by our finite number of tomography measurements, as the reconstruction protocol is Bayesian in character and uses a uniform prior.

\section{Outlook}

This work presents two significant achievements in quantum computing. First, we demonstrate fault-tolerant operation of logical qubits on a neutral atom platform, achieving up to a 15-fold reduction in error rates. Second, we conduct a small-scale practical material science simulation by determining the ground state of a family of single-impurity instances of the Anderson Impurity Model. These results underscore the critical role of quantum error correction codes as foundational elements for scalable, error-resilient quantum computing and advance the field toward quantum advantage.
 
Key next steps include scaling the number of logical qubits, implementing error correction codes with universal fault-tolerant gate sets, mid-circuit measurements, and multi-round error correction. These advancements will enable larger-scale applications with further reductions in error rates. Integrating GPUs and QPUs to optimize hybrid quantum-classical workloads will further enhance performance and scalability. Extending this framework to address a broader range of industrially relevant problems in materials science and chemical simulation will further validate progress and showcase the transformative potential of fault-tolerant quantum systems.
 
This work highlights neutral atom quantum computing as a leading platform, providing a strong foundation for the practical deployment of robust and scalable quantum technologies.

\section{Acknowledgments}
We thank Elica Kyoseva, Eric Anschuetz, Ikko Hamamura, Sam Stanwyck, and Yuri Alexeev for their insights.

This material is based upon work supported by the U.S. Department of Energy, Office of Science, Office of Advanced Scientific Computing Research, under Award Numbers DE-SC0021526 and DE-SC0025493. This material is based upon work supported by the U.S. Department of Energy, Office of Science, National Quantum Information Science Research Centers (Q-NEXT). Work on this manuscript is supported by Wellcome Leap as part of the Quantum for Bio (Q4Bio) Program. 

Disclaimer: This report was prepared as an account of work sponsored by an agency of the United States Government.  Neither the United States Government nor any agency thereof, nor any of their employees, makes any warranty, express or implied, or assumes any legal liability or responsibility for the accuracy, completeness, or usefulness of any information, apparatus, product, or process disclosed, or represents that its use would not infringe privately owned rights.  Reference herein to any specific commercial product, process, or service by trade name, trademark, manufacturer, or otherwise does not necessarily constitute or imply its endorsement, recommendation, or favoring by the United States Government or any agency thereof.  The views and opinions of authors expressed herein do not necessarily state or reflect those of the United States Government or any agency thereof.

\section{Author Contributions}

\begin{itemize}

\item W. C. Chung, D. C. Cole, G. T. Hickman, R. A. Jones, K. W. Kuper, M. T. Lichtman, D. Mason, T. W. Noel, A. G. Radnaev, and M. Saffman contributed to the design, construction, and operation of the experimental apparatus. 

\item M. J. Bedalov, M. Blakely, P. D. Buttler, W. C. Chung, D. C. Cole, P. Gokhale, G. T. Hickman, E. B. Jones, R. A. Jones, K. W. Kuper, M. T. Lichtman, D. Mason, N. A. Neff-Mallon, V. Omole, A. G. Radnaev, R. Rines, B. Thotakura, P. Khalate, JS. Kim, B. Heim, E. Shabtai, and A. K. Tucker contributed to the control system and software stack. 

\item W. C. Chung, D. C. Cole, P. Goiporia, P. Gokhale, E. B. Jones, K. W. Kuper, D. Mason, V. Omole, A. G. Radnaev, R. Rines, and B. Thotakura contributed to the collection and analysis of experimental data. 

\item F. T. Chong, D. C. Cole, P. Goiporia, P. Gokhale, E. B. Jones, K. W. Kuper, M. T. Lichtman, D. Mason, N. A. Neff-Mallon, V. Omole, R. Rines, M. H. Teo, and B. Thotakura contributed to theoretical analysis and simulation. 

\item The manuscript was written by F. T. Chong, W. C. Chung, P. Gokhale, E. B. Jones, K. W. Kuper, D. Mason, N. A. Neff-Mallon, V. Omole, A. G. Radnaev, R. Rines, M. Saffman, and M. H. Teo. 

\item The project was supervised by C. Carnahan, F. T. Chong, D. C. Cole, P. Gokhale, S. Lee, T. W. Noel, A. G. Radnaev, M. Saffman, and T. Tomesh.
\end{itemize}

\clearpage
\twocolumngrid

\bibliography{2_LQ_Assets/refs}
\bibliographystyle{unsrt}

\clearpage
\appendix

\section{Simulation Results} \label{app:simulation}

\subsection{Noise Model}
To better understand and substantiate our experiment results, we ran density matrix simulations of each circuit under a circuit-level noise model. We represent each atom as a five-level system $\{|0\rangle, |1\rangle, |0_\ell\rangle, |1_\ell\rangle, |L\rangle \}$, where $|0\rangle$ and $|1\rangle$ are the computational basis states, $|0_\ell\rangle$ ($|F = 3, m_F \neq 0\rangle$) and $|1_\ell\rangle$ ($|F = 4, m_F \neq 0\rangle$) are leakage states that are detected as $\{|0\rangle$ and $\{|1\rangle$ respectively, and $|L\rangle$ represents a loss state.

The noise model includes error mechanisms for each of the possible operations: state preparation, two-qubit gates (CZ), single-qubit gates (GR and $\text{R}_\text{Z}$), and measurement. These are summarized in Tab.~\ref{tab:sim_noise_model} and described in detail here. Note that this error model represents an estimate of system performance at one point in time, while not precisely corresponding to the fidelities in Fig. \ref{fig:Fig1}c and Fig. \ref{fig:perf_summary}

\begin{table} [h!]
\raggedright
\caption{Simulation Noise Model}
    \begin{tabular}{|c|m{0.8\columnwidth}|} \hline
         Operation& Error Mechanisms\\ \hline \hline
         
         State Prep & 
         \begin{minipage}[c][3cm]{0.8 \columnwidth} \raggedright
            \begin{itemize}
                 \item Initial state density matrix with $p=0.006$ and $q=0.046$: \\
                 $$
                 \begin{pmatrix}
                 \frac{p}{2} & 0 & 0 & 0 & 0 \\
                 0 & 1-p-q & 0 & 0 & 0 \\
                 0 & 0 & \frac{p}{2} & 0 & 0 \\
                 0 & 0 & 0 & q & 0 \\
                 0 & 0 & 0 & 0 & 0 \\
                 \end{pmatrix}
                 $$
                 \end{itemize}
         \end{minipage}
         \\ \hline
         
         CZ &
         \begin{minipage}[c][4cm][c]{0.8 \columnwidth}
             \begin{itemize}
                 \item Single qubit phase error with 0.35\% probability 
                 \item Spontaneous transitions with probabilities\\
                 $$
                 \begin{pmatrix}
                 17.4 & 185.0 & 4.9 & 165.4 & 0 \\
                 18.5 & 197.5 & 4.6 & 177.7 & 0 \\
                 31.1 & 420.3 & 45.8 & 1210.0 & 0 \\
                 42.1 & 590.2 & 52.9 & 1901.0 & 0 \\
                 0 & 5000 & 0 & 0 & 0 \\
                 \end{pmatrix} \times 10^{-6}
                 $$
             \end{itemize}
         \end{minipage}
         \\ \hline
         
         GR& 
         \begin{minipage}[c][1cm][c]{0.8 \columnwidth}
             \begin{itemize}
                 \item Static overrotation of 0.0345 radians
             \end{itemize}
        \end{minipage}
         \\ \hline
         
         $\text{R}_\text{Z}$& 
         \begin{minipage}[c][4cm][c]{0.8 \columnwidth}
             \begin{itemize}
                 \item Relative overrotation of $1.2\%$
                 \item Spontaneous transitions with probabilities\\
                 $$
                 \begin{pmatrix}
                 0 & 207.2 & 0 & 0 & 0 \\
                 0 & 223.1 & 0 & 0 & 0 \\
                 0 & 364.3 & 0 & 0 & 0 \\
                 0 & 524.0 & 0 & 0 & 0 \\
                 0 & 0 & 0 & 0 & 0 \\
                 \end{pmatrix} \times 10^{-6}
                 $$
             \end{itemize}
        \end{minipage}
         \\ \hline
         
         Measure& 
         \begin{minipage}[c][6.5cm][c]{0.8 \columnwidth}
             \begin{itemize}
                 \item Classification error:
                 $$\text{Prob}(|i\rangle) = \text{Tr}(\mathcal{P}_i \rho)$$ where for $\epsilon_0 = 0.004$ and $\epsilon_1 = 0.028$        
                 $$ \mathcal{P}_{0} = 
                 \begin{pmatrix}
                 1-\epsilon_0 & 0 & 0 & 0 & 0 \\
                 0 & \epsilon_1 & 0 & 0 & 0 \\
                 0 & 0 & 1-\epsilon_0 & 0 & 0 \\
                 0 & 0 & 0 & \epsilon_1 & 0 \\
                 0 & 0 & 0 & 0 & 0 \\
                 \end{pmatrix} 
                 $$
                 $$
                \mathcal{P}_{1} = 
                 \begin{pmatrix}
                 \epsilon_0 & 0 & 0 & 0 & 0 \\
                 0 & 1-\epsilon_1 & 0 & 0 & 0 \\
                 0 & 0 & \epsilon_0 & 0 & 0 \\
                 0 & 0 & 0 & 1-\epsilon_1 & 0 \\
                 0 & 0 & 0 & 0 & 0 \\
                 \end{pmatrix} 
                 $$
             \end{itemize}
        \end{minipage}
         \\  \hline
    \end{tabular}

    *All matrices expressed in the $\{|0\rangle, |1\rangle, |0_\ell\rangle, |1_\ell\rangle, |L\rangle \}$ basis
    
    \label{tab:sim_noise_model}
\end{table}

\begin{itemize}
    \item \textit{State Preparation-- } Due to imperfect optical repumping and microwave transfer, we assume some amount of leakage following the state preparation protocol. For each atom, we assume the total dark state leakage is $p$, split evenly between the states $|0\rangle$ and $|0_\ell\rangle$, and the total bright state leakage is $q$. Average values of $p$ $(0.006)$ and $q$ ($0.046)$ are estimated through measurements of the bright and dark state populations at each physical site following state preparation. As each qubit in the circuit is assumed to begin in $|0\rangle$, a GR $\pi$-pulse is further applied to this state as part of the state preparation step. 
    
    \item \textit{CZ Gate-- } Each CZ gate is subject to two error processes in our model: single-qubit phase errors and leakage/loss out of the computational basis.  The latter is calculated based on transition probabilities between the five levels of our system, including various decay pathways and atom loss (see \cite{radnaev2024universal} for details). The resulting transition probabilities are stated in Tab.~\ref{tab:sim_noise_model}. A single-qubit phase error probability of $0.35\%$ is found to produce simulated benchmarking results that match our observed depolarization. 
    
    \item \textit{GR Gate-- } The error in the GR gate is modeled as an absolute overrotation of 0.0345 radians per gate, such that an ideal GR gate with rotation axis $\phi$ and angle $\theta$ is simulated as GR($\phi$, $\theta + 0.0345$). This rotation error is chosen to match observed fidelities of $99.97\%$ in randomized benchmarking.
    
    \item \textit{$\text{R}_\text{Z}$ Gate-- } Each $\text{R}_\text{Z}$ gate is modeled as a combination of coherent rotation error and probabilistic leakage error.  The rotation error is assumed to be relative (an ideal $\text{R}_\text{Z}$ gate with rotation angle $\theta$ is simulated as $\text{R}_\text{Z}$($1.012 \theta$)), with a magnitude chosen to match randomized benchmarking.  The leakage error is calculated from a model of transition probabilities.  The total probability of leakage from $|1\rangle$ is $0.132\%$, and we estimate that approximately $15.7\%$ of this ends up in $|0\rangle$, $27.6\%$ in $|0_\ell\rangle$, $39.7\%$ in $|1_\ell\rangle$, and $16.9\%$ decays back to $|1\rangle$.
    
    \item \textit{Measurement-- } During non-destructive state-selective readout (NDSSR), there is some probability of incorrectly classifying bright states ($|F=4\rangle = \{|1\rangle,|1_\ell\rangle\}$) as dark states ($|F=3\rangle = \{|0\rangle,|0_\ell\rangle\}$), and vice versa. We attribute the measurement-induced bright-dark conversion to undesired depumping of bright states into dark states by the NDSSR excitation beam, and dark-bright conversion to small amounts of dark-to-bright repumping due to imperfect laser extinction. Informed by SPAM characterization measurements, we estimate the proportion of the bright population that is converted to dark states $\epsilon_1$ to be $0.028$ and the proportion of the dark population that is converted to bright states $\epsilon_0$ to be $0.004$.
\end{itemize}

This model is our best current approximation of the error mechanisms present within our system, informed by experimental characterization and benchmarking. However, system variability and imperfect estimates of each noise parameter, along with potential sources of error that are not captured within this model, mean that the model likely does not capture the full extent of the noise in the system. Some deviation between simulations and what we observe in experiment is thus expected. 

\subsection{Gottesman Benchmarking Simulations}

\begin{figure}[ht]
\centering
\includegraphics[width=1\columnwidth]{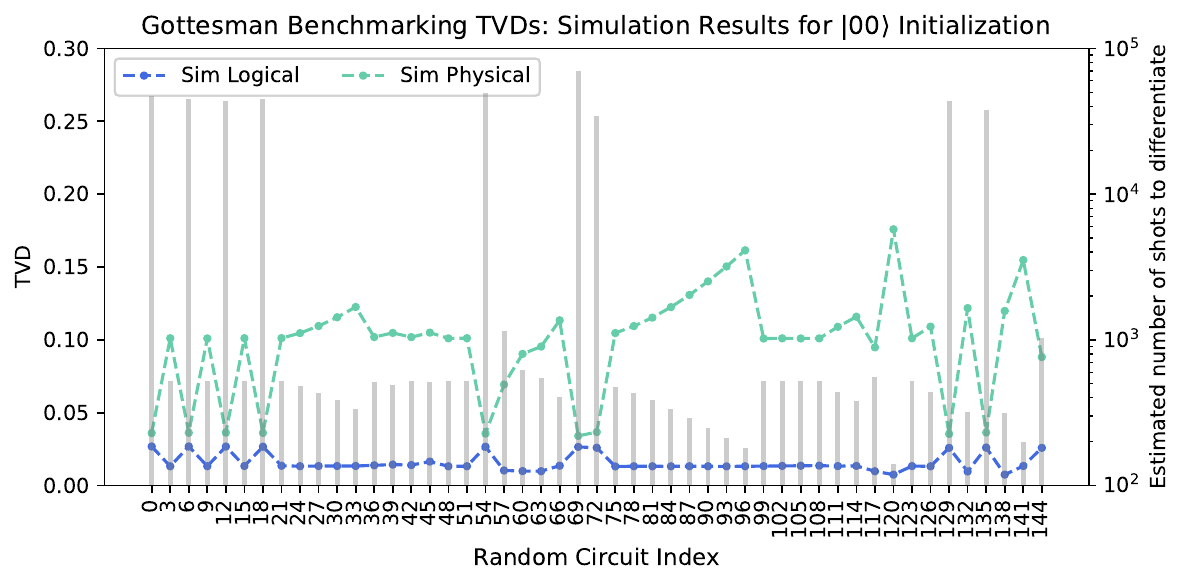}
\includegraphics[width=1\columnwidth]{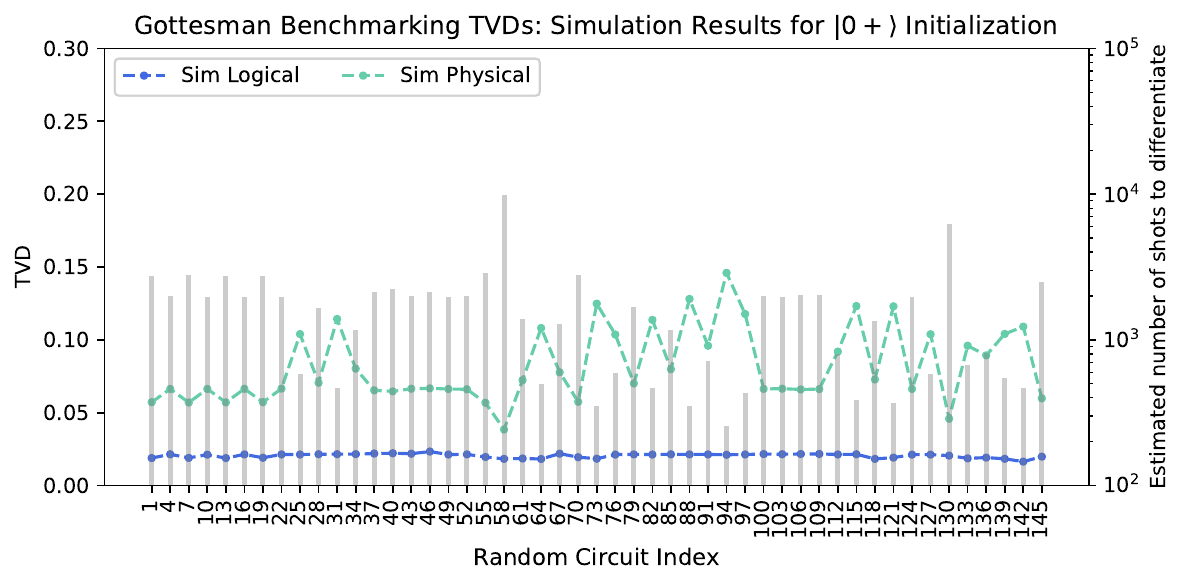}
\includegraphics[width=1\columnwidth]{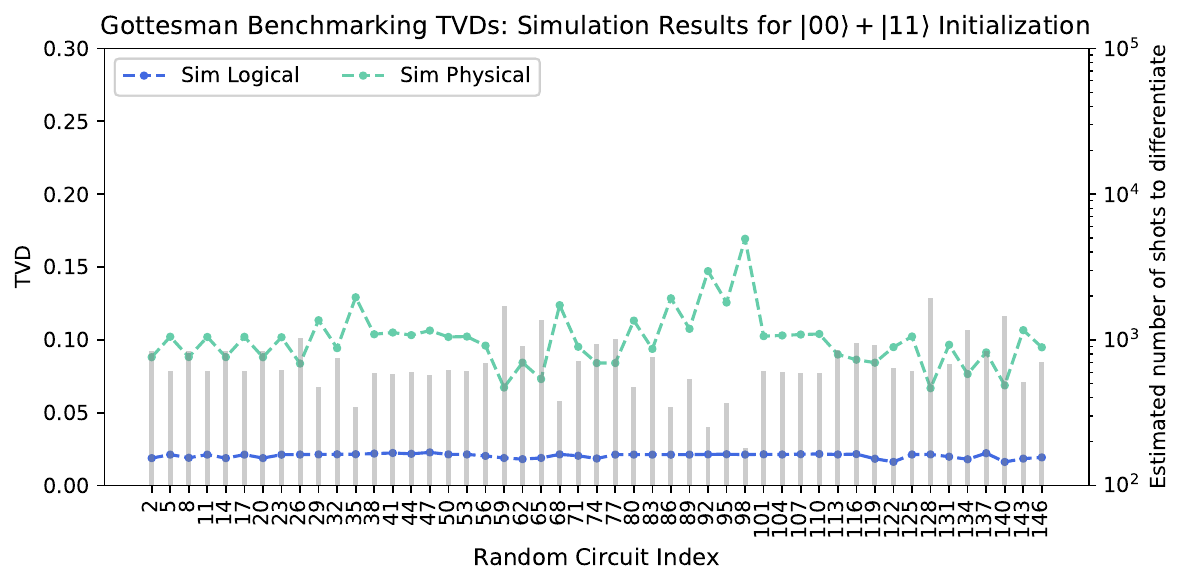}
\caption{Predicted TVDs for the Gottesman Benchmarking circuits. The lower blue line shows the TVDs for the simulated logical circuits, and the upper green line shows the TVDs for the simulated physical circuits. The gray bars show an estimate for the number of shots needed to differentiate between the logical and physical TVDs for each circuit.}
\label{fig:ap_golden_tvds}
\end{figure}

We apply our noise model to each of the Gottesman benchmarking circuits, the results of which are shown in Fig.~\ref{fig:ap_golden_tvds}. In general, we find that our simulations are in good agreement with experiment. We note that for circuits within the ranges 21-36 and 75-98, though we see qualitative agreement that the physical TVDs peak in those regions, quantitatively, there is an increasing deviation between simulations and experiment. As these circuits involve an increasing number of GR gates, this deviation may be due to an imperfect estimate of the coherent errors in our system. These uncaptured errors could build up in experiment, causing an increasing divergence from our results in simulation. Nonetheless, for the majority of circuits, our experiment results are consistent with simulation. 

In addition to verification of experiment results, these simulations also serve as a guide for working within a limited shot budget. For each circuit, we calculate an estimate for the number of shots that would be needed to differentiate the logical TVD from the physical TVD. In particular, we estimate that more than $10^4$ shots would be required to differentiate the circuits with indices \{0, 6, 12, 18, 54, 69, 72, 129, 135\}, which is all but one of the \texttt{PREP\_00} uniform distribution circuits. These simulations informed our decision to use an additional 7000 shots for the \texttt{PREP\_00} circuits with overlapping confidence intervals, and for the other initial states, validate that our inability to distinguish a performance advantage without additional shots is reasonable.  

\subsection{Anderson Impurity Model Simulations}
\begin{figure}
\centering
\includegraphics[width=1\columnwidth]{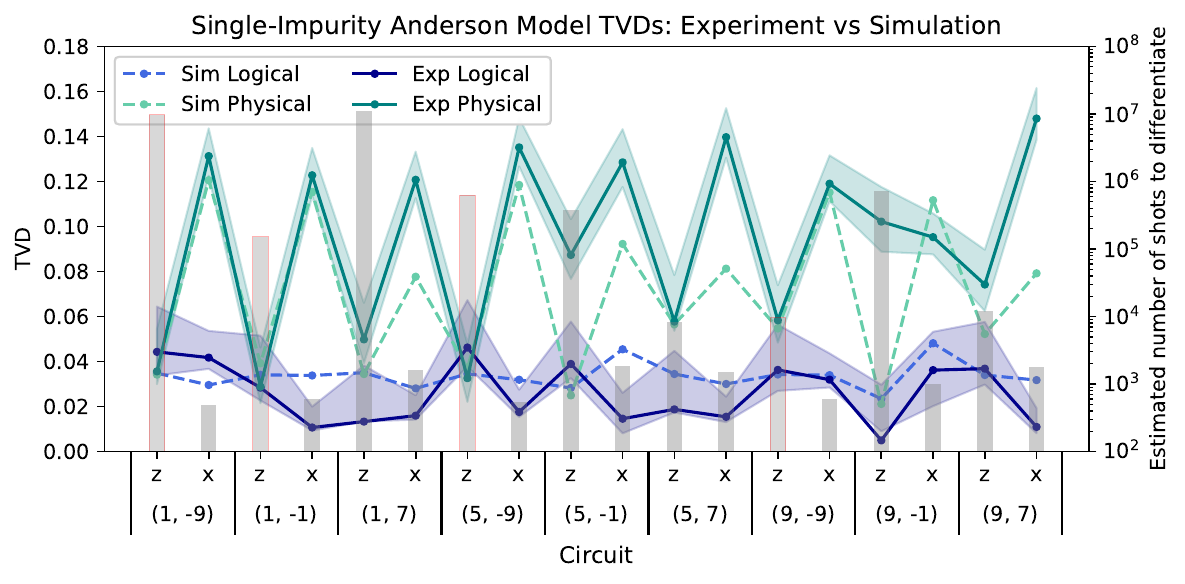}
\caption{Comparison of simulated and experimental TVD from the ideal output for each single-impurity Anderson Model circuit. The dashed (solid) lines represent simulation (experimental) results, and the blue (green) lines represent the logical (physical) circuits. The gray bars show an estimate for the number of shots needed to differentiate between the logical and physical TVDs for each circuit. Bars outlined in red denote circuits with overlapping logical and physical confidence intervals in experiment, showing that we estimate these circuits to require $10^4$-$10^7$ shots to demonstrate logical advantage. In general, the experiment results are consistent with what we find in simulation. }
\label{fig:ap_aim_tvds}
\end{figure}

\begin{figure} 
\centering
\includegraphics[width=1\columnwidth]{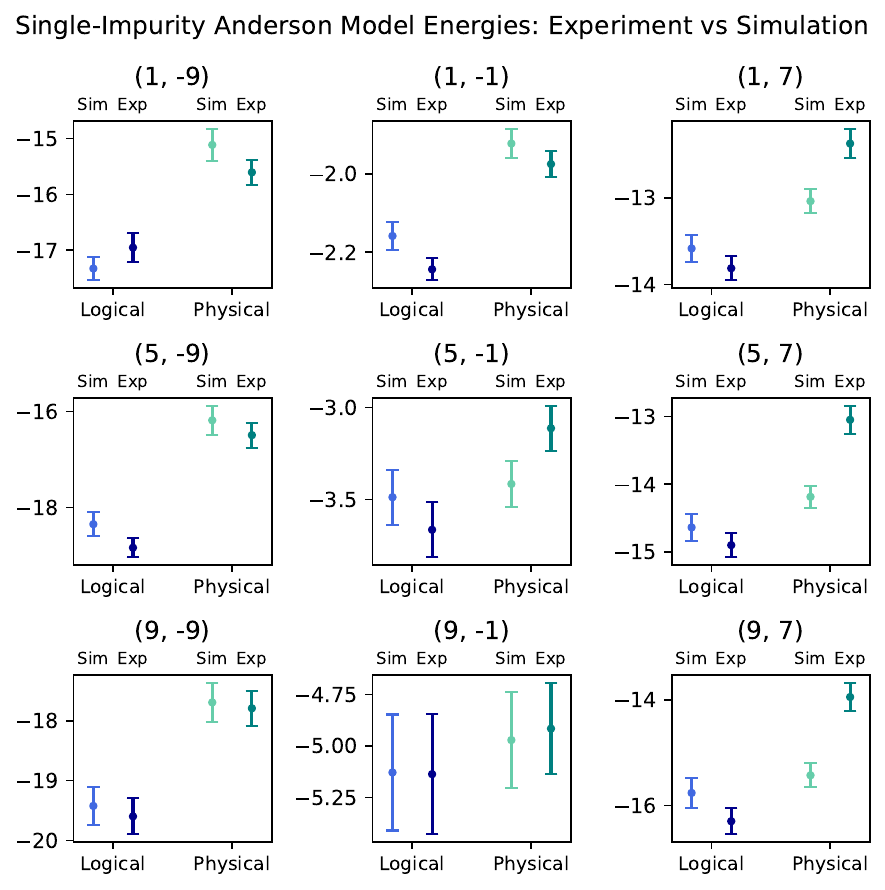}
\caption{Comparison of estimates of single-impurity Anderson Model ground state energies found in simulation and experiment at each pair of Hamiltonian parameters, U and V.}
\label{fig:ap_aim_energies}
\end{figure}

Our simulations of the single-impurity Anderson Model circuits are also consonant with what we find in experiment. Fig.~\ref{fig:ap_aim_tvds} shows a comparison of the TVDs that we find in experiment and in simulation, and Fig. \ref{fig:ap_aim_energies} compares the corresponding energy estimates.

Just as with the Gottesman circuits, we calculated an estimate of the number of shots needed to differentiate the simulated logical and physical TVDs of each circuit. For the circuits that had overlapping logical and physical confidence intervals in experiment -- (1,-9)-z, (1,-1)-z, (5,-9)-z, (9,-9)-z -- these estimates range from $10^4$ to $10^7$, far surpassing the actual shot budget allocated to these circuits. Again, these results broaden our understanding of the potential factors limiting our ability to differentiate logical performance from physical.

 Though the simulations are largely consistent with experiment, we observe particular divergence in the physical $U=5, V=-1$ circuits and the $V=7$ $x$-basis circuits. In these cases, the experiment results perform more poorly than the simulations predict, pointing again to potential error mechanisms that our current error model does not capture. Identifying currently unmodeled sources of error in the system is an ongoing area of investigation with which we hope to improve the accuracy of our simulations. 

\subsection{Using simulations to isolate the effects of individual noise parameters}
In our simulations of the Anderson Model circuits, there were several logical circuits that we predicted would not outperform their corresponding physical circuits. 

To better understand the effect of each individual noise parameter on the relative performance of the logical and physical circuits, we repeated the simulations of every circuit, tuning each of the error mechanisms listed in Tab.~\ref{tab:sim_noise_model} one at a time. Interestingly, we found that the GR overrotation parameter had a substantial effect on the fault tolerance of the circuits. As we scale the GR error, pseudothreshold-like behavior emerges: there is an error value above which the physical circuits outperform the logical circuits, and below which the reverse is true. 

\begin{figure} 
\centering
\includegraphics[width=1\columnwidth]{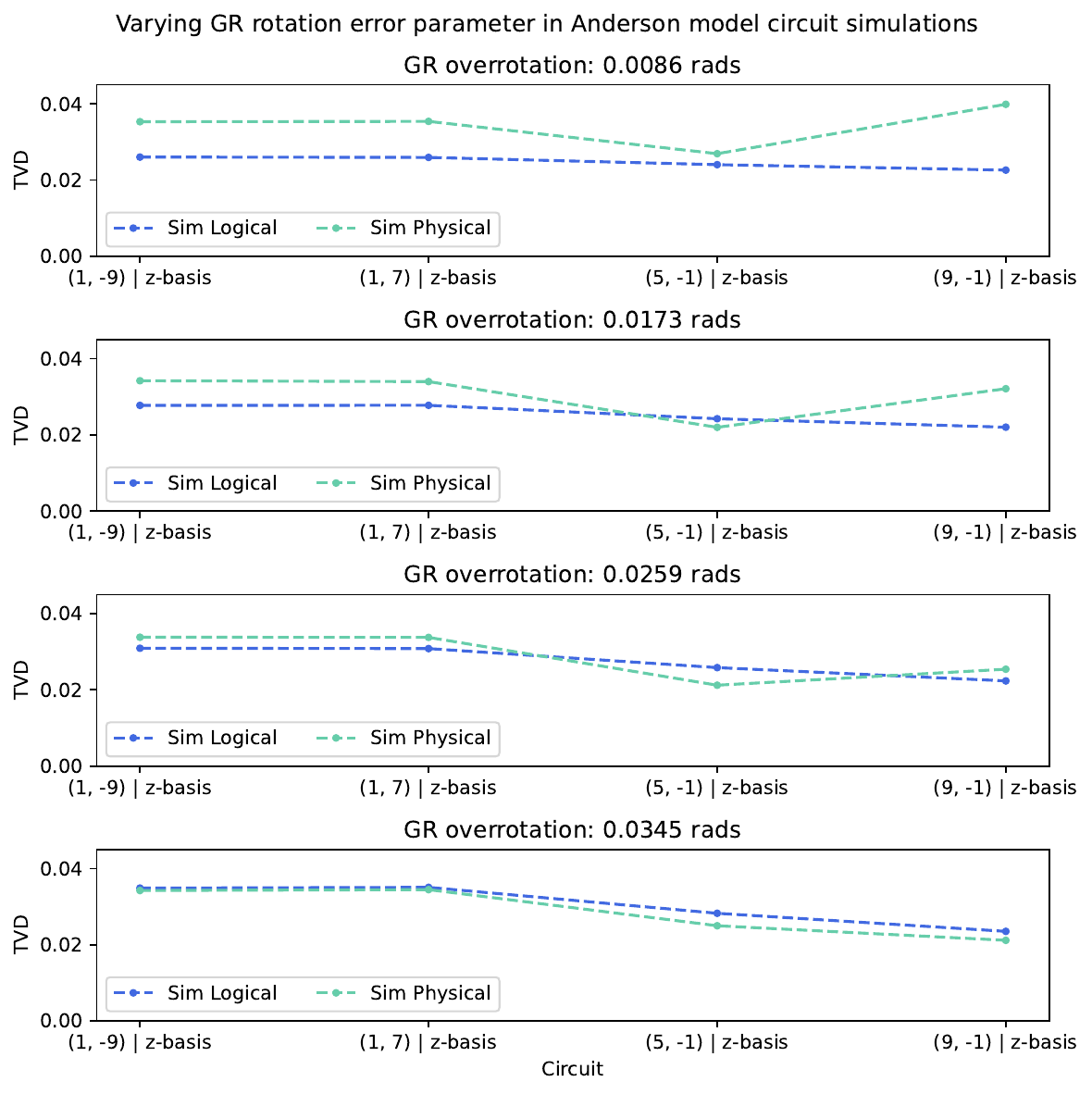}
\caption{Predicted TVDs of the circuits initially predicted to be non-fault tolerant as the GR overrotation error parameter is scaled from 8.6 mrad to 34.5 mrad.}
\label{fig:ap_aim_gr}
\end{figure}

Fig.~\ref{fig:ap_aim_gr} shows the simulated logical and physical TVDs for the circuits initially predicted to be non-fault-tolerant for different values of GR overrotation error. At $8.6$ mrad, or $0.25\times$ the GR overrotation error used in the original noise simulations, all of the logical circuits outperform their physical counterparts. As the GR overrotation error is scaled to $17.3$ mrad and $25.9$ mrad, or $0.5\times$ and $0.75\times$ the original value respectively, the logical and physical lines begin to cross. Finally, at 34.5 mrad, using the same noise model that was used to produce Figs.~\ref{fig:ap_aim_tvds} and~\ref{fig:ap_aim_energies}, all of the physical circuits begin to outperform the logical circuits. 

These results suggest that the optimization of the GR rotation gate could significantly enhance the relative performance of the [[4,2,2]] encoding in this system. As we improve our characterization of the errors in our system, our simulations can shed light on the sensitivity of system performance to individual error parameters.

\section{Experimental Details}
For the implementation details of single-qubit and two-qubit gates, see \cite{radnaev2024universal}. 
\begin{figure*}
    \centering
    \includegraphics[width=1\linewidth]{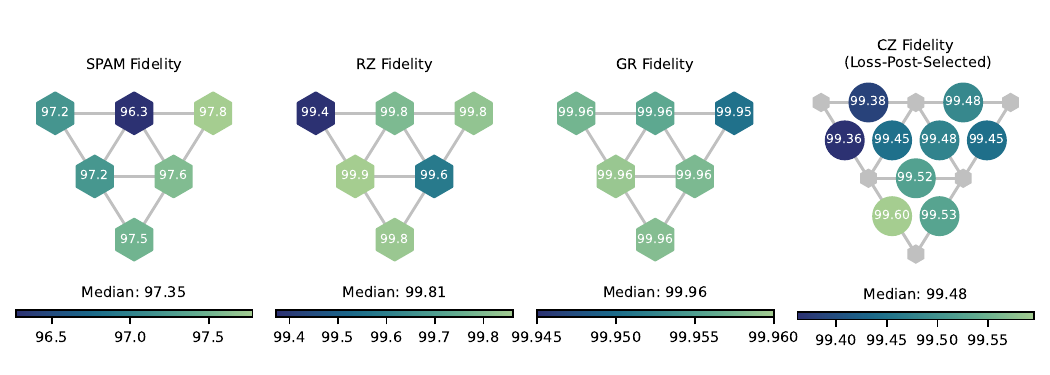}
    \caption{\textbf{System performance summary}. SPAM (State Prep and Measurement), RZ (local single-qubit $Z$ rotations), GR (global arbitrary single-qubit rotations), and CZ (controlled-Z) gate fidelities across the 6 qubits used in this work.  }
    \label{fig:enter-label}
\label{fig:perf_summary}
\end{figure*}
\subsection{Qubit Array generation}
Cesium atoms are trapped and cooled into a two-dimensional rectilinear grid of 1064~nm optical tweezers formed by two crossed AODs (AA Opto-Electronic DTSXY-400-1064), which are driven by multi-tone RF signals generated from a multi-channel signal generator (Quantum Machines OPX+). The first AOD is driven by a sum of five tones, with tone frequency spacing equal to  0.786~MHz, and the second AOD is driven by a sum of three tones, with tone frequency spacing equal to 1.361~MHz, such that the 1st diffraction order output of the crossed AODs is imaged into a $3\times 5$ grid with nominally $3\sqrt{3}$~$\mu\text{m}$ row spacing and 3~$\mu\text{m}$ column spacing. 

Prior to loading into the 1064~nm optical tweezers, cesium atoms are first stochastically loaded into a separate atom reservoir trap formed by crossings of 780~nm lines, which have nominally $3\sqrt{3}/2$~$\mu\text{m}$ row spacings and 3~$\mu\text{m}$ column spacings, and then rearranged such that they form a triangular lattice with 6~$\mu\text{m}$ spacing as shown in Fig.~\ref{fig:Fig1}. After rearrangement into the triangular lattice pattern, they are transferred to the 1064~nm traps. 

\subsection{State preparation and measurement}
Once cesium atoms are deterministically loaded into the 1064~nm traps, they are first cooled by polarization gradient cooling (PGC) and then further cooled by Raman sideband cooling (RSC). Cesium atoms are left mostly in the $|F=4, m_F=4\rangle$ hyperfine level after RSC and are transferred to the upper qubit level $|F=4, m_F=0\rangle = |1\rangle$ by a series of composite microwave pulses. The microwave transitions used are $|F=4, m_F=4\rangle \rightarrow |F=3, m_F=3\rangle$, $|F=3, m_F=3\rangle \rightarrow |F=4, m_F=2\rangle$, $|F=4, m_F=2\rangle \rightarrow |F=3, m_F=1\rangle$, and $|F=3, m_F=1\rangle \rightarrow |F=4, m_F=0\rangle$ and each transition is performed with a Knill pulse sequence for robustness against Rabi frequency and microwave detuning errors.

Both RSC and the microwave state transfer are performed while the magnetic bias is parallel to the grid plane, with 10.1~G bias.  After the end of the microwave state transfer, the magnetic bias is rotated to the direction of the optical axis of the individual addressing beams (normal of the grid plane), with 11.1~G bias in a cubic spline ramp. The cesium atoms are then transferred to the lower qubit level $|F=3, m_F=0\rangle = |0\rangle$ by a microwave pi-pulse, which marks the end of qubit preparation before quantum circuit execution.

After the execution of a quantum circuit, qubit state is readout by a non-destructive state-selective readout (NDSSR) sequence described in \cite{radnaev2024universal}, which acquires bright camera signal from $|F=4\rangle$ hyperfine levels and dark signal from $|F=3\rangle$ levels. The state readout is subsequently followed by an occupancy readout for checking atom occupancy loss. We note that the NDSSR sequence used here is slightly modified from the one described in \cite{radnaev2024universal} as the two NDSSR illuminations are no longer simultaneously on but instead alternate with zero overlap and higher saturation parameters (similar to the imaging beam illumination sequence in \cite{su2024fastimaging}). Despite the modification, the overall performance has not changed, with 1\% state-average atom loss rate per site and roughly 3\% bright-to-dark-state depumping error during state measurement.

We note that both PGC and RSC return hotter temperatures in the $3\times 5$ grid than they did for $1 \times 2$ grid with 6~$\mu\text{m}$ spacing used in \cite{radnaev2024universal} to achieve 99.35(4)\,\% $CZ$ fidelity with individual optical addressing. For RSC, the final average temperature of the array increased from approximately 2.6~$\mu\text{K}$ to 8.5~$\mu\text{K}$. Also, the oscillation contrast of the global rotation  between $|0\rangle$ and $1\rangle$ is inhomogeneous in the new grid, whereas the oscillation contrast between $|F=4, m_F=0\rangle = |1\rangle$ and $|F=3, m_F=1\rangle$ appears much less inhomogeneous prior to the magnetic field rotation. We suspect that both cooling performance degradation and qubit state preparation fidelity degradation are due to intensity modulation sidebands at each optical tweezer spot, which can arise due to a combination of RF and acousto-optical nonlinearities in the RF generator + AOD system. We observed that microwave Rabi oscillation contrasts for transitions involved in the microwave state transfer became more homogeneous only after reducing the trap depth adiabatically prior to the state transfer sequence. The same reduced trap depth is maintained during the magnetic field rotation and then adiabatically raised back after the field rotation is complete, but we did not observe corresponding improvement in the qubit microwave transition oscillation contrast after the rotation. However, reducing the duration of the bias field rotation after the microwave state transfer (initially from 4~ms to 1.5~ms) improved the oscillation contrasts on some sites and reduced the minimum-maximum difference of SPAM fidelities across the array from 3.8\% to 1.5\% shown in Fig.~\ref{fig:perf_summary}.

\section{TVD Error} \label{app:tvd_error}
To estimate the uncertainty envelope for the TVD for a particular circuit result, the TVD distribution is estimated from the Dirichlet distribution corresponding to the circuit result counts. We approximate the integral of this distribution by sampling the Dirichlet distribution 10,000 times and calculating the TVD for each sampled point. We then calculate the narrowest band of TVD values that includes 68\% of the observed samples and report this interval as the uncertainty envelope. The occurrence of a measured TVD less than the lower bound of the uncertainty envelope is consistent with a system where the posterior TVD distribution is dominated by projection noise.

\clearpage

\section{Gottesman Protocol Circuits} \label{app:gottesman_circuits}

\begin{table}[h!]
    \centering
    \begin{tabular}{c|l}
0--2 & HH \\ \hline
3--5 & HH HH \\ \hline
6--8 & HH HH HH \\ \hline
9--11 & HH HH HH HH \\ \hline
12--14 & HH HH HH HH HH \\ \hline
15--17 & HH HH HH HH HH HH \\ \hline
18--20 & HH HH HH HH HH HH HH \\ \hline
21--23 & HH HH HH HH HH HH HH HH \\ \hline
24--26 & XC IZ \\ \hline
27--29 & XC IZ XC IZ \\ \hline
30--32 & XC IZ XC IZ XC IZ \\ \hline
33--35 & XC IZ XC IZ XC IZ XC IZ \\ \hline
36--38 & IX IX CZ \\ \hline
39--41 & IX IX CZ IX IX CZ \\ \hline
42--44 & HH IZ HH IX \\ \hline
45--47 & HH IZ HH IX HH IZ HH IX \\ \hline
48--50 & (no-op) \\ \hline
51--53 & IZ \\ \hline
54--56 & ZI HH \\ \hline
57--59 & XI ZI IX \\ \hline
60--62 & XC IX IZ IZ \\ \hline
63--65 & CX ZI XI XC CZ \\ \hline
66--68 & XC CZ IZ XI XI XC \\ \hline
69--71 & ZI IX CX ZI HH CZ IZ \\ \hline
72--74 & CX XC CZ ZI XI XC XC HH \\ \hline
75--77 & XC \\ \hline
78--80 & XC XC \\ \hline
81--83 & XC XC XC \\ \hline
84--86 & XC XC XC XC \\ \hline
87--89 & XC XC XC XC XC \\ \hline
90--92 & XC XC XC XC XC XC \\ \hline
93--95 & XC XC XC XC XC XC XC \\ \hline
96--98 & XC XC XC XC XC XC XC XC \\ \hline
99--101 & ZI ZI \\ \hline
102--104 & ZI ZI ZI ZI \\ \hline
105--107 & ZI ZI ZI ZI ZI ZI \\ \hline
108--110 & ZI ZI ZI ZI ZI ZI ZI ZI \\ \hline
111--113 & CZ XC CX \\ \hline
114--116 & CZ XC CX CZ XC CX \\ \hline
117--119 & XC CX IX IZ \\ \hline
120--122 & XC CX IX IZ XC CX IX IZ \\ \hline
123--125 & ZI \\ \hline
126--128 & CX XC \\ \hline
129--131 & IZ ZI HH \\ \hline
132--134 & IZ XC IX XC \\ \hline
135--137 & XI IX IX CZ HH \\ \hline
138--140 & IX IZ IZ XC IZ IZ \\ \hline
141--143 & XI CZ ZI CZ CX CZ XC \\ \hline
144--146 & HH HH IX IZ HH IZ ZI IX \\ \hline
    \end{tabular}
    \caption{All 147 circuits benchmarked under Gottesman protocol with maximum depth $T = 8$, $r = 2$ randomization rounds, and maximum period $p = 4$. The first 74 circuits correspond to the first randomization round and the remaining 73 correspond to the second randomization round. Within each group of three, the executions were ordered as PREP\_00, PREP\_0+, PREP\_BELL.}
    \label{tab:my_label}
\end{table}


\newcolumntype{O}[1]{>{\centering\let\newline\\\arraybackslash\hspace{0pt}}p{#1}}
\newcolumntype{C}[1]{>{\centering\scriptsize\let\newline\\\arraybackslash\hspace{0pt}}p{#1}}
\begin{table*}
\caption{\centering Fault-tolerant circuits for state preparation, logical gate, and measurement operations for the [[4, 2, 2]] code. The compiled circuits used Superstaq for implementations that minimize the number of global rotation gates and single-site $R_Z$ rotation gates.
For the logical \texttt{HH} gate, two different compiled sequences are presented: one designed to leave the flag qubit undisturbed (used for circuits initialized with \texttt{PREP\_00}), and one used for circuits with no flag qubit (i.e. those initialized with \texttt{PREP\_0+} and \texttt{PREP\_BELL}). Both also make use of a virtual global-$Z$ operation, implemented by flipping the sign of all subsequent GR gates.
All other compiled sequences for the logical gates leave the flag qubit undisturbed. \label{tab:ft_operations}}
\hspace*{-0.6cm}
\centering
\begin{tabular}{|O{1.8cm}|C{1.7cm}|C{2.8cm}|C{12.0cm}|}
\hline
Operation & \normalsize Unencoded & \normalsize FT Encoded & \normalsize Compiled \\

\hline \hline
\rowcolor{green!10}
{\texttt{PREP\_00\newline(q0, q1)}} &
\begin{minipage}{\linewidth}
\Qcircuit @C=1em @R=1em {
& \lstick{\ket{0}} & \qw \\
& \lstick{\ket{0}} & \qw \\
}
\end{minipage}
&
\begin{minipage}{\linewidth}
\hspace{8pt}
\Qcircuit @C=0.2em @R=0.5em {
&\lstick{\ket{0}} & \qw  & \qw & \targ & \qw & \ctrl{4} & \qw \\
&\lstick{\ket{0}} & \gate{H} & \ctrl{1} & \ctrl{-1}  & \qw & \qw & \qw \\
&\lstick{\ket{0}} & \qw & \targ & \ctrl{1} & \qw & \qw  & \qw\\
&\lstick{\ket{0}} & \qw & \qw & \targ & \ctrl{1} & \qw & \qw \\
&\lstick{\ket{0}} & \qw & \qw & \qw & \targ & \targ & \meter \\
}

\end{minipage}
&
\begin{minipage}{\linewidth}
\hspace{4pt}
\vspace{2pt}
\Qcircuit @R=0em @C=0.5em {
&\lstick{\ket{0}}&\multigate{4}{\GR{\frac{\pi}{2}}{\frac{\pi}{2}}} \qw&\qw&\ctrl{1}&\gate{\RZ{\rzfrac{-\pi}{2}}}&\multigate{4}{\GR{\frac{\pi}{2}}{\frac{\pi}{2}}} \qw&\qw&\qw&\multigate{4}{\GR{\frac{\pi}{2}}{\pi}} \qw&\ctrl{4}&\qw&\multigate{4}{\GR{\frac{\pi}{2}}{\frac{\pi}{2}}} \qw&\qw&\multigate{4}{\GR{\frac{-\pi}{2}}{\frac{\pi}{2}}} \qw&\qw \\
&\lstick{\ket{0}}&\ghost{\GR{\frac{\pi}{2}}{\frac{\pi}{2}}} \qw&\ctrl{1}&\control \qw&\qw&\ghost{\GR{\frac{\pi}{2}}{\frac{\pi}{2}}} \qw&\qw&\gate{\RZ{\rzfrac{-\pi}{2}}}&\ghost{\GR{\frac{\pi}{2}}{\pi}} \qw&\qw&\qw&\ghost{\GR{\frac{\pi}{2}}{\frac{\pi}{2}}} \qw&\qw&\ghost{\GR{\frac{-\pi}{2}}{\frac{\pi}{2}}} \qw&\qw \\
&\lstick{\ket{0}}&\ghost{\GR{\frac{\pi}{2}}{\frac{\pi}{2}}} \qw&\control \qw&\qw&\qw&\ghost{\GR{\frac{\pi}{2}}{\frac{\pi}{2}}} \qw&\ctrl{1}&\qw&\ghost{\GR{\frac{\pi}{2}}{\pi}} \qw&\qw&\qw&\ghost{\GR{\frac{\pi}{2}}{\frac{\pi}{2}}} \qw&\gate{\RZ{\rzfrac{\pi}{2}}}&\ghost{\GR{\frac{-\pi}{2}}{\frac{\pi}{2}}} \qw&\qw \\
&\lstick{\ket{0}}&\ghost{\GR{\frac{\pi}{2}}{\frac{\pi}{2}}} \qw&\qw&\qw&\gate{\RZ{\rzfrac{\pi}{2}}}&\ghost{\GR{\frac{\pi}{2}}{\frac{\pi}{2}}} \qw&\control \qw&\qw&\ghost{\GR{\frac{\pi}{2}}{\pi}} \qw&\qw&\ctrl{1}&\ghost{\GR{\frac{\pi}{2}}{\frac{\pi}{2}}} \qw&\qw&\ghost{\GR{\frac{-\pi}{2}}{\frac{\pi}{2}}} \qw&\qw \\
&\lstick{\ket{0}}&\ghost{\GR{\frac{\pi}{2}}{\frac{\pi}{2}}} \qw&\qw&\qw&\qw&\ghost{\GR{\frac{\pi}{2}}{\frac{\pi}{2}}} \qw&\qw&\gate{\RZ{\rzfrac{-\pi}{2}}}&\ghost{\GR{\frac{\pi}{2}}{\pi}} \qw&\control \qw&\control \qw&\ghost{\GR{\frac{\pi}{2}}{\frac{\pi}{2}}} \qw&\gate{\RZ{\rzfrac{\pi}{2}}}&\ghost{\GR{\frac{-\pi}{2}}{\frac{\pi}{2}}} \qw&\qw
\\
}
\vspace{2pt}

\end{minipage}
\\

\hline

\rowcolor{green!10}
\texttt{PREP\_0+\newline(q0, q1)} &
\begin{minipage}{\linewidth}
\Qcircuit @C=1em @R=1em {
& \lstick{\ket{0}} & \qw & \qw \\
& \lstick{\ket{0}} & \gate{H} & \qw
}
\end{minipage}
&
\begin{minipage}{\linewidth}
\Qcircuit @C=.5em @R=0.5em {
& \lstick{\ket{0}} & \gate{H} & \ctrl{1} & \qw \\
& \lstick{\ket{0}} & \qw & \targ & \qw \\
& \lstick{\ket{0}} & \gate{H} & \ctrl{1} & \qw \\
& \lstick{\ket{0}} & \qw & \targ & \qw
}
\end{minipage}
&
\begin{minipage}{\linewidth}
\vspace{2pt}
\Qcircuit @R=0.4em @C=.5em {
&\lstick{\ket{0}}&\multigate{3}{\GR{\frac{\pi}{2}}{\frac{\pi}{2}}} \qw&\ctrl{1}&\multigate{3}{\GR{\frac{\pi}{2}}{0}} \qw&\gate{\RZ{\rzfrac{\pi}{2}}}&\multigate{3}{\GR{\frac{\pi}{2}}{\frac{-\pi}{2}}} \qw&\qw \\
&\lstick{\ket{0}}&\ghost{\GR{\frac{\pi}{2}}{\frac{\pi}{2}}} \qw&\control \qw&\ghost{\GR{\frac{\pi}{2}}{0}} \qw&\qw&\ghost{\GR{\frac{\pi}{2}}{\frac{-\pi}{2}}} \qw&\qw \\
&\lstick{\ket{0}}&\ghost{\GR{\frac{\pi}{2}}{\frac{\pi}{2}}} \qw&\ctrl{1}&\ghost{\GR{\frac{\pi}{2}}{0}} \qw&\gate{\RZ{\rzfrac{\pi}{2}}}&\ghost{\GR{\frac{\pi}{2}}{\frac{-\pi}{2}}} \qw&\qw \\
&\lstick{\ket{0}}&\ghost{\GR{\frac{\pi}{2}}{\frac{\pi}{2}}} \qw&\control \qw&\ghost{\GR{\frac{\pi}{2}}{0}} \qw&\qw&\ghost{\GR{\frac{\pi}{2}}{\frac{-\pi}{2}}} \qw&\qw \\
}

\end{minipage}

\\

\hline

\rowcolor{green!10}
\texttt{PREP\_BELL\newline(q0, q1)}
& 
\begin{minipage}{\linewidth}
\hspace{6pt}
\Qcircuit @C=.5em @R=1em {
& \lstick{\ket{0}} & \gate{H} & \ctrl{1} & \qw \\
& \lstick{\ket{0}} & \qw & \targ & \qw 
}
\end{minipage}
&
\begin{minipage}{\linewidth}
\Qcircuit @C=0.5em @R=0.4em {
& \lstick{\ket{0}} & \gate{H} & \qw & \ctrl{3} & \qw \\
& \lstick{\ket{0}} & \gate{H} & \ctrl{1} & \qw & \qw \\
& \lstick{\ket{0}} & \qw & \targ & \qw & \qw \\
& \lstick{\ket{0}} & \qw & \qw & \targ & \qw
}
\end{minipage} 
&
\begin{minipage}{\linewidth}
\vspace{2pt}
\Qcircuit @R=0.4em @C=0.5em {
&\lstick{\ket{0}}&\multigate{3}{\GR{\frac{\pi}{2}}{\frac{\pi}{2}}} \qw&\ctrl{3}&\qw&\multigate{3}{\GR{\frac{\pi}{2}}{0}} \qw&\gate{\RZ{\rzfrac{\pi}{2}}}&\multigate{3}{\GR{\frac{\pi}{2}}{\frac{-\pi}{2}}} \qw&\qw \\
&\lstick{\ket{0}}&\ghost{\GR{\frac{\pi}{2}}{\frac{\pi}{2}}} \qw&\qw&\ctrl{1}&\ghost{\GR{\frac{\pi}{2}}{0}} \qw&\gate{\RZ{\rzfrac{\pi}{2}}}&\ghost{\GR{\frac{\pi}{2}}{\frac{-\pi}{2}}} \qw&\qw \\
&\lstick{\ket{0}}&\ghost{\GR{\frac{\pi}{2}}{\frac{\pi}{2}}} \qw&\qw&\control \qw&\ghost{\GR{\frac{\pi}{2}}{0}} \qw&\qw&\ghost{\GR{\frac{\pi}{2}}{\frac{-\pi}{2}}} \qw&\qw \\
&\lstick{\ket{0}}&\ghost{\GR{\frac{\pi}{2}}{\frac{\pi}{2}}} \qw&\control \qw&\qw&\ghost{\GR{\frac{\pi}{2}}{0}} \qw&\qw&\ghost{\GR{\frac{\pi}{2}}{\frac{-\pi}{2}}} \qw&\qw \\
}
\end{minipage}

\\

\hline \hline

\rowcolor{blue!10}
\texttt{X(q0)} &
\begin{minipage}{\linewidth}
\Qcircuit @C=1em @R=1em {
& \gate{X} & \qw \\
& \qw & \qw
}
\end{minipage}
&
\begin{minipage}{\linewidth}
\Qcircuit @C=1em @R=0.5em {
& \gate{X} & \qw \\
& \qw & \qw \\
& \gate{X} & \qw \\
& \qw & \qw
}
\end{minipage} 
&
\begin{minipage}{\linewidth}
\vspace{2pt}
\Qcircuit @R=0.4em @C=0.5em {
&\multigate{3}{\GR{\frac{-\pi}{2}}{\frac{\pi}{2}}} \qw&\gate{\RZ{\pi}}&\multigate{3}{\GR{\frac{\pi}{2}}{\frac{\pi}{2}}} \qw&\qw \\
&\ghost{\GR{\frac{-\pi}{2}}{\frac{\pi}{2}}} \qw&\qw&\ghost{\GR{\frac{\pi}{2}}{\frac{\pi}{2}}} \qw&\qw \\
&\ghost{\GR{\frac{-\pi}{2}}{\frac{\pi}{2}}} \qw&\gate{\RZ{\pi}}&\ghost{\GR{\frac{\pi}{2}}{\frac{\pi}{2}}} \qw&\qw \\
&\ghost{\GR{\frac{-\pi}{2}}{\frac{\pi}{2}}} \qw&\qw&\ghost{\GR{\frac{\pi}{2}}{\frac{\pi}{2}}} \qw&\qw \\
}
\vspace{2pt}
\end{minipage}
\\

\hline

\rowcolor{blue!10}
\texttt{Z(q0)} &
\begin{minipage}{\linewidth}
\Qcircuit @C=1em @R=1em {
& \gate{Z} & \qw\\
& \qw & \qw
}
\end{minipage}
&
\begin{minipage}{\linewidth}
\Qcircuit @C=1em @R=0.5em {
& \gate{Z} & \qw \\
& \gate{Z} & \qw \\
& \qw & \qw \\
& \qw & \qw
}
\end{minipage} 
&
\begin{minipage}{\linewidth}
\Qcircuit @R=0.5em @C=1em {
&\gate{\RZ{\pi}}&\qw \\
&\gate{\RZ{\pi}}&\qw \\
&\qw&\qw \\
&\qw&\qw \\
}
\vspace{2pt}

\end{minipage}
\\

\hline

\rowcolor{blue!10}
\texttt{X(q1)} &
\begin{minipage}{\linewidth}
\Qcircuit @C=1em @R=1em {
& \qw & \qw\\
& \gate{X} & \qw
}
\end{minipage}
&
\begin{minipage}{\linewidth}
\Qcircuit @C=1em @R=0.5em {
& \gate{X} & \qw \\
& \qw & \qw \\
& \gate{X} & \qw \\
& \qw & \qw
}

\end{minipage}
&
\begin{minipage}{\linewidth}
\vspace{2pt}
\Qcircuit @R=0.5em @C=0.5em {
&\multigate{3}{\GR{\frac{-\pi}{2}}{\frac{\pi}{2}}} \qw&\gate{\RZ{\pi}}&\multigate{3}{\GR{\frac{\pi}{2}}{\frac{\pi}{2}}} \qw&\qw \\
&\ghost{\GR{\frac{-\pi}{2}}{\frac{\pi}{2}}} \qw&\gate{\RZ{\pi}}&\ghost{\GR{\frac{\pi}{2}}{\frac{\pi}{2}}} \qw&\qw \\
&\ghost{\GR{\frac{-\pi}{2}}{\frac{\pi}{2}}} \qw&\qw&\ghost{\GR{\frac{\pi}{2}}{\frac{\pi}{2}}} \qw&\qw \\
&\ghost{\GR{\frac{-\pi}{2}}{\frac{\pi}{2}}} \qw&\qw&\ghost{\GR{\frac{\pi}{2}}{\frac{\pi}{2}}} \qw&\qw \\
}

\end{minipage}

 \\

\hline

\rowcolor{blue!10}
\texttt{Z(q1)} &
\begin{minipage}{\linewidth}
\Qcircuit @C=1em @R=1em {
& \qw & \qw\\
& \gate{Z} & \qw
}
\end{minipage}
&
\begin{minipage}{\linewidth}
\Qcircuit @C=1em @R=0.5em {
& \gate{Z} & \qw \\
& \qw & \qw \\
& \gate{Z} & \qw \\
& \qw & \qw
}
\end{minipage} 
&
\begin{minipage}{\linewidth}
\Qcircuit @R=0.4em @C=1em {
&\gate{\RZ{\pi}}&\qw \\
&\qw&\qw \\
&\gate{\RZ{\pi}}&\qw \\
&\qw&\qw \\
}
\vspace{2pt}
\end{minipage}
\\

\hline \hline

\rowcolor{blue!10}
\texttt{CZ(q0, q1)} &
\begin{minipage}{\linewidth}
\Qcircuit @C=1.0em @R=1.0em {
& \ctrl{1} & \qw \\
& \control \qw & \qw
}
\end{minipage}
&
\begin{minipage}{\linewidth}
\Qcircuit @R=0.4em @C=1em {
&\gate{S}&\qw \\
&\gate{S^\dagger}&\qw \\
&\gate{S^\dagger}&\qw \\
&\gate{S}&\qw \\
}

\end{minipage} 
&
\begin{minipage}{\linewidth}
\Qcircuit @R=.4em @C=1em {
&\gate{\RZ{\rzfrac{ \pi}{2}}}&\qw \\
&\gate{\RZ{\rzfrac{-\pi}{2}}}&\qw \\
&\gate{\RZ{\rzfrac{-\pi}{2}}}&\qw \\
&\gate{\RZ{\rzfrac{ \pi}{2}}}&\qw \\
}
\vspace{2pt}
\end{minipage} 
\\

\hline

\rowcolor{blue!10}
\texttt{CX(q0, q1)} &
\begin{minipage}{\linewidth}
\Qcircuit @C=1em @R=1em {
& \ctrl{1} & \qw \\
& \targ \qw & \qw
}
\end{minipage}
&
\begin{minipage}{\linewidth}
\vspace{4pt}
\Qcircuit @C=1em @R=1em {
& \qswap & \qw \\
& \qswap \qwx & \qw \\
& \qw & \qw \\
& \qw & \qw
}
\vspace{4pt}
\end{minipage} 
&
\begin{minipage}{\linewidth}
\Qcircuit @C=1em @R=1em {
& \qw & \link{1}{-1} & \qw \\
& \qw & \link{-1}{-1} & \qw \\
& \qw & \qw & \qw \\
& \qw & \qw & \qw
}
\end{minipage} 
\\

\hline

\rowcolor{blue!10}
\texttt{CX(q1, q0)} &
\begin{minipage}{\linewidth}
\Qcircuit @C=1em @R=1em {
& \targ & \qw \\
& \ctrl{-1} \qw & \qw
}
\end{minipage}
&
\begin{minipage}{\linewidth}
\vspace{4pt}
\Qcircuit @C=1em @R=1em {
& \qswap \qwx[2] & \qw \\
& \qw & \qw \\
& \qswap & \qw \\
& \qw & \qw \\
}
\vspace{4pt}
\end{minipage}
&
\begin{minipage}{\linewidth}
\Qcircuit @C=1em @R=1em {
& \qw & \link{2}{-1} & \qw \\
& \qw & \qw & \qw \\
& \qw & \link{-2}{-1} & \qw \\
& \qw & \qw & \qw
}
\end{minipage} 
\\

\hline

\rowcolor{blue!10}

\texttt{HH(q0, q1)} &
\begin{minipage}{\linewidth}
\Qcircuit @C=1em @R=1em {
& \gate{H} & \qw \\
& \gate{H} & \qw
}
\end{minipage}
&
\begin{minipage}{\linewidth}
\Qcircuit @C=1em @R=0.4em {
& \gate{H} & \qw & \qw \\
& \gate{H} & \qswap & \qw \\
& \gate{H} & \qswap \qwx & \qw \\
& \gate{H} & \qw & \qw
}
\end{minipage} 
&
\begin{minipage}{\linewidth}
\begin{tabular}{c}
\texttt{PREP\_00}:\hspace{0.1em}
\Qcircuit @R=0.4em @C=0.5em {
&\multigate{4}{\GR{\frac{-\pi}{4}}{\frac{\pi}{2}}} \qw&\qw            &\multigate{4}{\GR{\frac{-\pi}{4}}{\frac{\pi}{2}}}\qw&\multigate{4}{\VZ{}} \qw&\qw&\qw&\qw \\
&       \ghost{\GR{\frac{-\pi}{4}}{\frac{\pi}{2}}} \qw&\qw            &       \ghost{\GR{\frac{-\pi}{4}}{\frac{\pi}{2}}}\qw&       \ghost{\VZ{}} \qw&\qw & \link{1}{-1}&\qw \\
&       \ghost{\GR{\frac{-\pi}{4}}{\frac{\pi}{2}}} \qw&\qw            &       \ghost{\GR{\frac{-\pi}{4}}{\frac{\pi}{2}}}\qw&       \ghost{\VZ{}} \qw&\qw & \link{-1}{-1} &\qw \\
&       \ghost{\GR{\frac{-\pi}{4}}{\frac{\pi}{2}}} \qw&\qw            &       \ghost{\GR{\frac{-\pi}{4}}{\frac{\pi}{2}}}\qw&       \ghost{\VZ{}} \qw&\qw&\qw&\qw \\
&       \ghost{\GR{\frac{-\pi}{4}}{\frac{\pi}{2}}} \qw&\gate{\RZ{\pi}}&       \ghost{\GR{\frac{-\pi}{4}}{\frac{\pi}{2}}}\qw&       \ghost{\VZ{}} \qw&\qw&\qw&\qw\\
}

\hspace{1em}\texttt{PREP\_0+}, \texttt{PREP\_BELL}:\hspace{0.1em}
\Qcircuit @R=0.4em @C=0.5em {
&\multigate{3}{\GR{\frac{\pi}{2}}{\frac{-\pi}{2}}} \qw&\multigate{3}{\VZ{}} \qw&\qw&\qw&\qw \\
&       \ghost{\GR{\frac{\pi}{2}}{\frac{-\pi}{2}}} \qw&       \ghost{\VZ{}} \qw&\qw& \link{1}{-1} &\qw \\
&       \ghost{\GR{\frac{\pi}{2}}{\frac{-\pi}{2}}} \qw&       \ghost{\VZ{}} \qw&\qw& \link{-1}{-1} &\qw \\
&       \ghost{\GR{\frac{\pi}{2}}{\frac{-\pi}{2}}} \qw&       \ghost{\VZ{}} \qw&\qw&\qw&\qw\\
}
\end{tabular}
\end{minipage}
\\

\hline \hline

\rowcolor{red!10}
\texttt{MEAS\newline(q0, q1)} &
\begin{minipage}{\linewidth}
\Qcircuit @C=1em @R=1em {
& \meter \\
& \meter
}
\end{minipage}
&
\begin{minipage}{\linewidth}
\Qcircuit @C=0.5em @R=0.4em {
&\qw& \meter \\
&\qw& \meter \\
&\qw& \meter \\
&\qw& \meter
} 
\end{minipage} 
&
\begin{minipage}{\linewidth}
\Qcircuit @C=0.5em @R=0.4em {
&\qw& \meter \\
&\qw& \meter \\
&\qw& \meter \\
&\qw& \meter
}
\end{minipage} 
\\

\hline



\end{tabular}
\end{table*}

\end{document}